\documentclass[structabstract]{aa}  

\usepackage{graphicx}
\usepackage{txfonts}

\begin{document}

   \title{Line formation in AGB atmospheres including velocity effects}

   \subtitle{Molecular line profile variations of long period variables}

   \author{W.~Nowotny\inst{1}   \and
           S.~H\"ofner\inst{2}  \and
           B.~Aringer\inst{1} 
           }

   \institute{University of Vienna, Department of Astronomy, 
              T\"urkenschanzstra{\ss}e 17, A-1180 Wien, Austria\\
              \email{walter.nowotny@univie.ac.at}
         \and
              Department of Physics and Astronomy, 
              Division of Astronomy and Space Physics,
              Uppsala University,
              Box 515, SE-75120 Uppsala, Sweden\\
             }

   \date{Received; accepted}

\titlerunning{ Line formation in AGB atmospheres including velocity effects}
\authorrunning{W. Nowotny et al.}

 
\abstract
{Towards the end of the evolutionary stage of the Asymptotic Giant Branch (AGB) the atmospheres of evolved red giants are considerably influenced by radial pulsations of the stellar interiors and developing stellar winds. The resulting complex velocity fields severely affect molecular line profiles (shapes, time-dependent shifts in wavelength, multiple components) observable in near-infrared spectra of long period variables. \textrm{Time-series high-resolution spectroscopy allows us to probe the atmospheric kinematics and} thereby study the mass loss process.}
{With the help of model calculations the complex line formation process in AGB atmospheres was explored with the focus on velocity effects. Furthermore, we aimed for atmospheric models which are able to quantitatively reproduce line profile variations found in observed spectra of pulsating late-type giants.}
{Models describing pulsation-enhanced dust-driven winds were used to compute synthetic spectra under the assumptions of chemical equilibrium and LTE. For this purpose, we used molecular data from line lists for the considered species and solved the radiative transfer in spherical geometry including the effects of velocity fields. Radial velocities (RV) derived from Doppler-shifted (components of) synthetic line profiles provide information on the gas velocities in the line-forming region of the spectral features. In addition, we made use of radial optical depth distributions to give estimates for the layers where lines are formed and to illustrate the effects of velocities in the line formation process.}
{Assuming uniform gas velocities for all depth points of an atmospheric model we estimated the conversion factor between gas velocities and measured RVs to $p$\,=\,$u_{\rm gas}$\,/\,$RV$\,$\approx$1.2--1.5. On the basis of dynamic model atmospheres and by applying our spectral synthesis codes we investigated in detail the finding that various molecular features in AGB spectra originate at different geometrical depths of the very extended atmospheres of these stars. We show that the models are able to quantitatively reproduce the characteristic line profile variations of lines sampling the deep photosphere (CO $\Delta v$\,=\,3, CN) of Mira variables and the corresponding discontinuous, S-shaped RV curve. The global velocity fields (traced by different features) of typical long-period variables are also realistically reproduced. \textrm{Possible reasons for discrepancies concerning other modelling results (e.g. CO $\Delta v$\,=\,2 lines) are outlined.} In addition, we present a model showing variations of CO $\Delta v$\,=\,3 line profiles comparable to observed spectra of semiregular variables and discuss that the non-occurence of line doubling in these objects may be due to a density effect.}
{The results of our line profile modelling are another \textrm{indication}  that the dynamic models studied here \textrm{are approaching a realistic representation of the outer layers of AGB stars with or without mass loss.}
}


   \maketitle

\section{Introduction}\label{s:intro}

Stars on the Asymptotic Giant Branch (AGB) represent objects of low to intermediate main sequence mass ($\approx$0.8--8\,$M_{\odot}$) in a late evolutionary phase. While they exhibit low effective temperatures ($<$3500\,K), their luminosities can reach values of up to a few 10$^4\,L_{\odot}$ at the tip of the AGB, placing them in the upper right corner of the Hertzsprung-Russell diagram. Compared to the atmospheres of most other types of stars the outer layers of these evolved red giants have remarkable properties (see e.g. the review by Gustafsson \& H\"ofner \cite{GustH04}, below GH04). 

In the cool and very extended atmospheres (extensions of the same order as the radii of the stars; up to a few 100\,$R_{\odot}$), molecules can form. Their large number of internal degrees of freedom results in a plethora of spectral lines. Thus, molecules significantly affect the spectral appearance of late-type giants at visual and infrared wavelengths (IR; e.g. Lan\c{c}on \& Wood \cite{LancW00}, Gautschy-Loidl et al. \cite{GaHJH04}, GH04, Aringer et al. \cite{AGNML09}). 

On the upper part of the AGB, the stars become instable to strong radial pulsations. This leads to a pronounced variability of the emitted flux with amplitudes of up to several magnitudes in the visual (e.g. Lattanzio \& Wood \cite{LattW04}). Since the variations occur on long time scales of a few 10 to several 100 days, pulsating AGB stars are often referred to as \textit{long period variables} (LPVs). In the past, different types of LPVs were empirically classified according to the regularity of the light change and the visual light amplitude: Mira variables (regular, $\Delta V$$>$2.5$^{\rm m}$), semiregular variables (SRVs, poor regularity, $\Delta V$$<$2.5$^{\rm m}$), and irregular variables (irregular, $\Delta V$$<$1--2$^{\rm m}$). Major advances in our understanding of the pulsation of AGB stars were achieved by exploiting the data sets of surveys for microlensing events (MACHO, OGLE, EROS), which produced a substantial number of high-quality lightcurves for red variables as a by-product. According to the pioneering work in this field by Wood et al. (\cite{WAAAA99}) and Wood (\cite{Wood00}), and to a number of subsequent studies (e.g. Lebzelter et al. \cite{LebSM02c}, Ita et al. \cite{ITMNN04a}, \cite{ITMNN04b}) it is probably more adequate to characterise LPVs according to their pulsation mode than to their light change in the visual as it was done historically. From observational studies during recent years (see e.g. Lattanzio \& Wood \cite{LattW04}) it appears that stars start to pulsate (as SRVs) in the second/third overtone mode (corresponding to sequence A in Fig.\,1 of Wood \cite{Wood00}) and switch then to the first overtone mode (sequence B). Light amplitudes are increasing while the stars evolve and finally become Miras. In this stadium they pulsate in the fundamental mode (sequence C) and show highly periodic light changes. Observational evidence for this evolution scenario was found for LPVs in the globular cluster 47\,Tuc by Lebzelter et al. (\cite{LWHJF05b}) and Lebzelter \& Wood (\cite{LebzW05d}).

The pulsating stellar interior of an AGB star severely influences the outer layers. The atmospheric structure is periodically modulated, and in the wake of the emerging shock waves dust condensation can take place. Radiation pressure on the newly formed dust grains (at least in the C-rich case, cf. Sect.\,\ref{s:DMAgenrem}) leads to the development of a rather slow (terminal velocities of max. 30\,km\,s$^{-1}$) but dense stellar wind with high mass loss rates (from a few 10$^{-8}$ up to 10$^{-4}$\,$M_{\odot}$\,yr$^{-1}$; e.g. Olofsson \cite{Olofs04}). 

As a consequence of these dynamic processes -- pulsation and mass loss -- the atmospheres of evolved AGB stars eventually become even more extended than non-pulsating red giants in earlier evolutionary stages. The resulting atmospheric structure strongly deviates from a hydrostatic configuration and shows temporal variations on global and local scales (Sect.\,\ref{s:modelling}). The complex, non-monotonic velocity fields with relative macroscopic motions of the order of 10\,km\,s$^{-1}$ have substantial influence on the shapes of individual spectral lines (Doppler effect). Observational studies have demonstrated that time series high-resolution spectroscopy in the near IR (where AGB stars are bright and well observable) is a valuable tool to study atmospheric kinematics throughout the outer layers of pulsating and mass-losing red giants (e.g. Hinkle et al. \cite{HinHR82}, from now on HHR82, or Alvarez et al. \cite{AJPGF00}). Radial velocities (RV) derived from Doppler-shifts of various spectral lines provide clues on the gas velocities in the line-forming regions of the respective features. A detailed review on studies of line profile variations for AGB stars (observations and modelling) can be found for example in Nowotny (\cite{Nowot05}, below N05).

In two previous papers (Nowotny et al. \cite{NAHGW05}+\cite{NoLHH05}, from now on Paper\,I and Paper\,II, respectively) we investigated whether observed variations of line profiles can be comprehended with state-of-the-art dynamic model atmospheres. We were able to show that the used models allow to qualitatively reproduce the behaviour of spectral lines originating in different regions of the extended atmospheres. The work presented here can be regarded as an extension of the previous two papers about line profile modelling. The aim is to shed light on the intricate line formation process within the atmospheres of evolved red giants with an emphasis on the velocity effects (Sects.\,\ref{s:lineformation}, \ref{s:velocityeffects}, \ref{s:simulating} and \ref{s:when-linedoubling}). In addition, we report on our efforts to achieve realistic models, which are able to reproduce line profile variations and the derived RVs even quantitatively (Sects.\,\ref{s:realistic} and \ref{s:COdv3SRVs}).

\section{Model atmospheres and spectral synthesis}\label{s:modelling}

\subsection{General remarks}\label{s:DMAgenrem}

Modelling the cool and very extended atmospheres of evolved AGB stars remains challenging due to the intricate interaction of different complicated phenomena (convection, pulsation, radiation, molecular and dust formation/absorption, acceleration of winds). Dynamic model atmospheres are constructed to simulate and understand the physical processes (e.g. mass loss) occuring in the outer layers of AGB stars. In particular they are needed if one is interested in reproducing the complex and temporally varying atmospheric structures that form the basis for radiative transfer calculations, which allows us to simulate observational results (spectra, photometry, etc.).

For our line profile modelling we used dynamic model atmospheres as described in detail by H\"ofner et al. (\cite{HoGAJ03}; DMA3), Gautschy-Loidl et al. (\cite{GaHJH04}; DMA4), N05 or Papers\,I+II. These models represent the scenario of pulsation-enhanced dust-driven winds (cf. Sects.\,4.7+4.8 of GH04). They provide a consistent and realistic description from the deep and dust-free photosphere (dominated by the pulsation of the stellar interior) out to the dust-forming layers and beyond to the stellar wind region at the inner circumstellar envelope (characterised by the cool, steady outflow). This is accomplished by a combined and self-consistent solution of hydrodynamics, frequency-dependent radiative transfer and a detailed time-dependent treatment of dust formation and evolution.

\begin{table}
\begin{center}
\caption{Characteristics of the dynamic atmospheric models for pulsating, C-rich AGB stars (for a detailed description see text and DMA3) used for the modelling.}
\begin{tabular}{ll|ccc}
\hline
\hline
\multicolumn{2}{l|}{Model:}& W & S & M \\
\hline
$L_\star$&[$L_{\odot}$]&7000&10\,000&7000\\
$M_\star$&[$M_{\odot}$]&1.0&1.0&1.5\\
$T_\star$&[K]&2800&2600&2600\\
$[$Fe/H$]$&[dex]&0.0&0.0&0.0\\
C/O&\textit{by number}&1.4&1.4&1.4\\
\hline
$R_\star$&[$R_{\odot}$]&355&493&412\\
&[\textit{AU}]&1.65&2.29&1.92\\
log $g_\star$& &--0.66&--0.94&--0.61\\
\hline
$P$&[d]&390&490&490\\
$\Delta u_{\rm p}$&[km\,s$^{-1}$]&2&4&6\\
$f_{\rm L}$& &1.0&2.0&1.5\\
$\Delta m_{\rm bol}$&[mag]&0.21&0.86&1.07\\
\hline
$\langle\dot M\rangle$&[$M_{\odot}\
$yr$^{-1}$]&--&4.3$\cdot$10$^{-6}$&2.5$\cdot$10$^{-6}$\\
$\langle u \rangle$&[km\,s$^{-1}$]&--&15&7.5\\
$\langle f_c$$\rangle$& &--&0.28&0.40\\
\hline
\end{tabular}
\label{t:dmaparameters}
\end{center}
Notes: Listed are (i) parameters of the hydrostatic initial model, (ii) quantities derivable from these parameters, (iii) attributes of the inner boundary (piston) used to simulate the pulsating stellar interior as well as the resulting bolometric amplitude $\Delta m_{\rm bol}$, and (iv) properties of the resulting wind. The notation was adopted from previous papers (DMA3, DMA4): $P$, $\Delta u_{\rm p}$ -- period and velocity amplitude of the piston at the inner boundary; $f_L$ -- free parameter to adjust the luminosity amplitude at the inner boundary; $\langle\dot M\rangle$, $\langle u \rangle$ -- mean mass loss rate and outflow velocity at the outer boundary; $\langle f_c$$\rangle$ -- mean degree of condensation of carbon into dust at the outer boundary. The radial coordinates in this work are plotted in units of the corresponding stellar radii $R_\star$ of the hydrostatic initial models, calculated from their luminosities $L_\star$ and temperatures $T_\star$ (as given in the table) via the relation $L_\star$\,=\,4$\pi$$R_\star^2$\,$\sigma$$T_\star^4$.
\end{table}

As a result of deep-reaching convection (dredge-up), nucleo-synthesis products can be mixed up from the stellar interior of AGB stars, resulting in a metamorphosis of the molecular chemistry of the whole atmosphere (e.g. Busso et al. \cite{BusGW99}, Herwig \cite{Herwi05}). The most important product of all the nuclearly processed material mixed up is carbon $^{12}$C. As a consequence, the stars can turn from oxygen-rich (C/O$<$1) to carbon-rich (C/O$>$1) during the late AGB phase. The resulting so-called \textit{carbon stars} (C~stars) can be found close to the tip of the AGB in observed colour-magnitude diagrams (e.g. Nowotny et al. \cite{NoKOS03}). The drastic change of the atmospheric chemical composition is not only relevant for the observable spectral type of the star (changing from M to C), it is also crucial for the formation of circumstellar dust. 

Observational studies revealed a rich mineralogy in the dusty outflows of O-rich objects (e.g. Molster \& Waters \cite{MolsW03} and references therein). Unfortunately, the dust formation process is theoretically not fully understood. Many physical and chemical details of the process are still not clear, which would be necessary for a fully consistent numerical treatment. In addition to the lack of a grain formation theory, there is an ongoing debate concerning the underlying physics of the driving mechanism (Woitke \cite{Woitk06b}+\cite{Woitk07}, H\"ofner \cite{Hoefn07}, H\"ofner \& Andersen \cite{HoefA07}). A potential solution was recently suggested by H\"ofner (\cite{Hoefn08}).

\begin{figure}
\resizebox{\hsize}{!}{\includegraphics[clip]{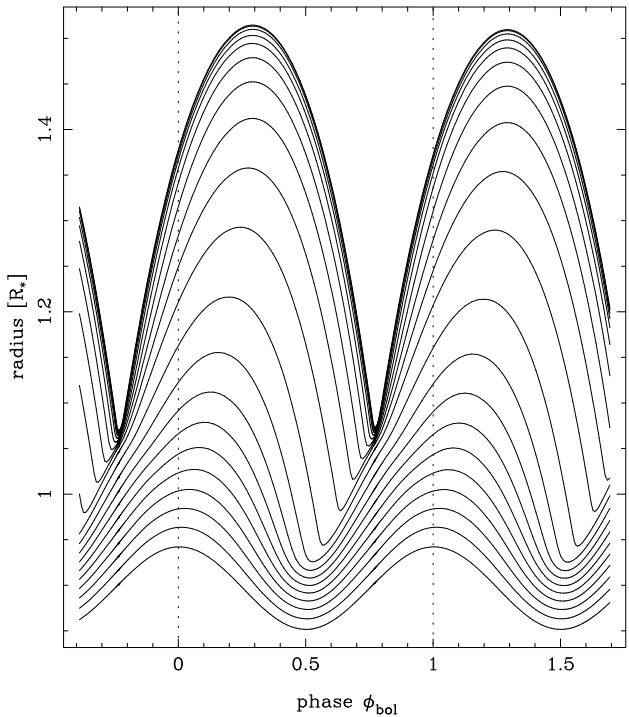}}
\caption{Movement of mass shells with time at different depths of the dust-free
model~W, which exhibits no mass loss and shows a strictly periodic behaviour
for all layers. Note the different scales on the radius-ordinates compared to Fig.\,\ref{f:massenschalenM}. The shown trajectories represent the points of the adaptive grid at a selected instance of time (higher density of points at the location of shocks) and their evolution with time (cf. the caption of Fig.\,2 in DMA3 for further explanations).}
\label{f:massenschalenW}
\end{figure}

\begin{figure}
\resizebox{\hsize}{!}{\includegraphics[clip]{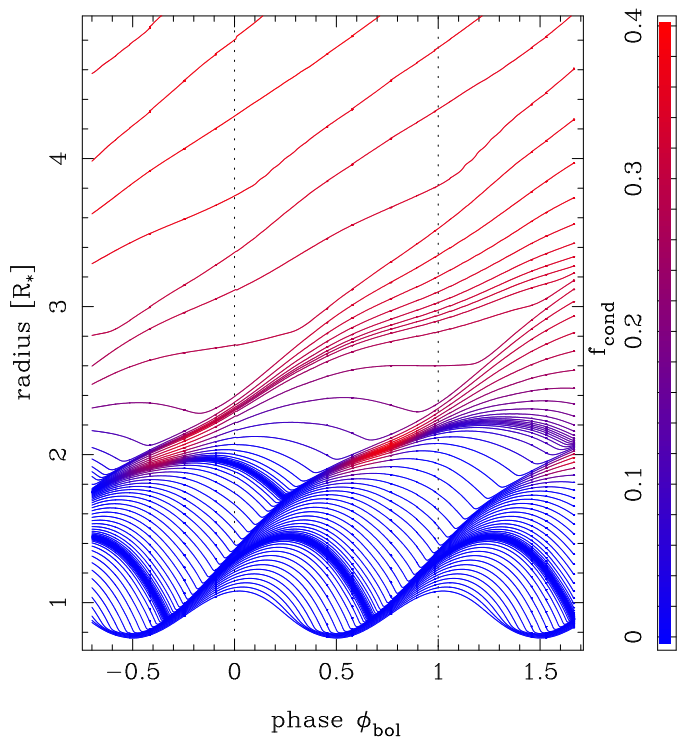}}
\caption{Same plot as Fig.\,\ref{f:massenschalenW} for model~M, representing the scenario of a pulsation-enhanced dust-driven wind. The plot illustrates the different regions within the atmosphere of a typical mass-losing LPV. The innermost, dust-free layers below $\approx$2\,$R_{\star}$ are subject to strictly regular motions caused by the pulsating interior (shock fronts). The dust-forming region (colour-coded is the degree of dust condensation $f_{\rm c}$) at $\approx$2--3\,$R_{\star}$ where the stellar wind is triggered represents dynamically a transition region with moderate velocities, not necessarily periodic. A continuous outflow is found from $\approx$4\,$R_{\star}$ outwards, where the dust-driven wind is decisive from the dynamic point of view.}
\label{f:massenschalenM}
\end{figure}

Things are quite different for C-rich stars, where we find a rather simple composition of the circumstellar dust. A very limited variety of dust species were identified by their spectral features (e.g. Molster \& Waters \cite{MolsW03}), as for example SiC (prominent feature at $\approx$11$\mu$m) or MgS (broad emission band around 30$\mu$m). The most important species is amorphous carbon dust, though. Not producing any distinctive spectral feature, grains of carbon dust represent the dominating condensate and play a crucial role from the dynamic point of view (mass loss process). Amorphous carbon fulfills the relevant criteria for a catalyst of dust-driven winds: (i) made up of abundant elements, (ii) simple and efficient formation process, (iii) refractory, i.e. stable at high temperatures ($\approx$1500\,K), (iv) large radiative cross section around 1$\mu$m in order to absorb momentum. Moreover, the formation and evolution of amorphous carbon dust grains can be treated numerically in a consistent way by using moment equations as described in Gail \& Sedlmayr (\cite{GailS88}) and Gauger et al. (\cite{GauGS90}). As a consequence of this, atmospheric models for C-rich AGB stars with dusty outflows were quite successfully calculated and applied in the past by different groups (see the overviews given by Woitke  \cite{Woitk03} or H\"ofner et al. \cite{HGANH04}), among them the models used in this work.

\subsection{Models for atmospheres and winds}\label{s:DMAsused}

\begin{figure*}
\centering
\includegraphics[width=17cm,clip]{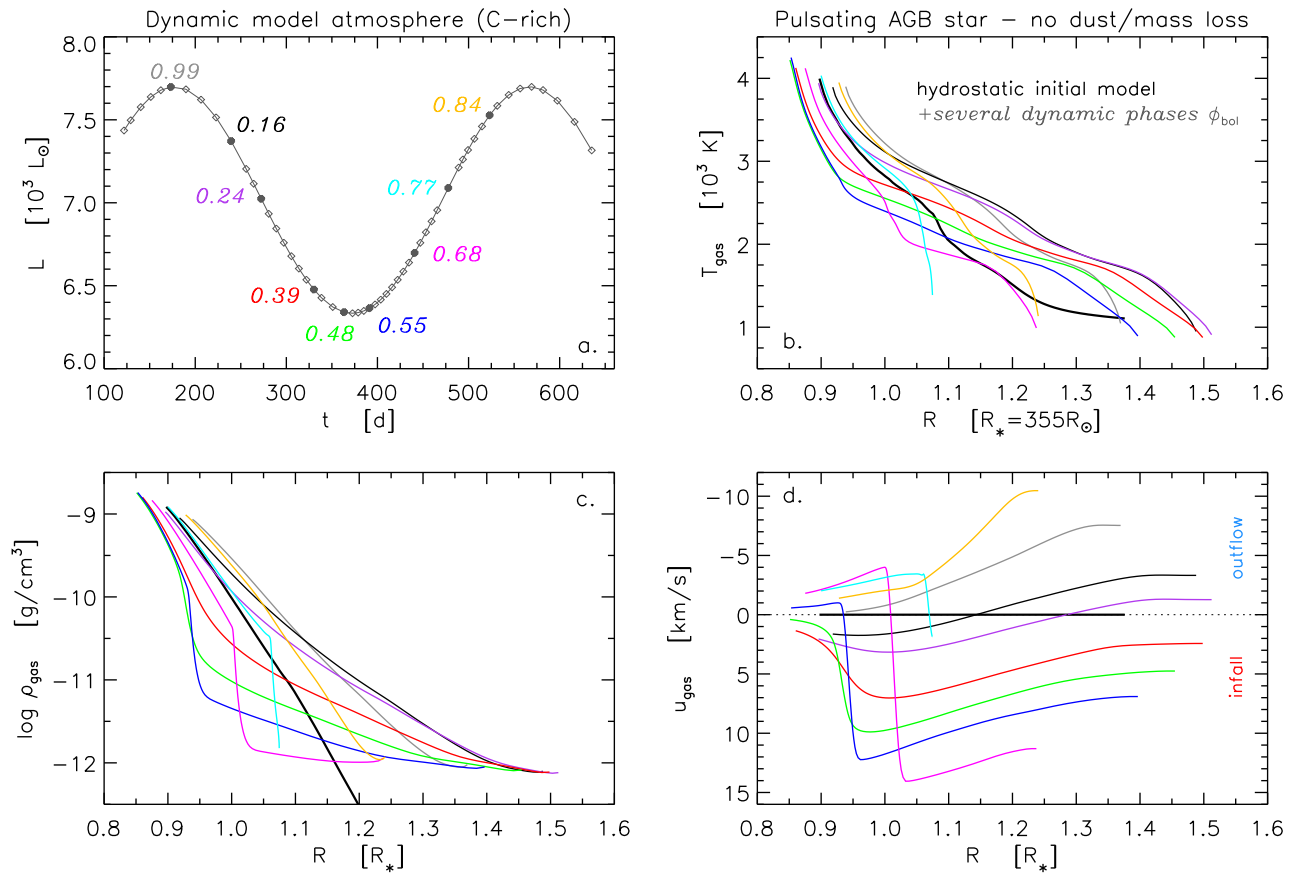}
\caption{Characteristic properties of model~W. Plotted in panel (a) is the bolometric lightcurve resulting from the variable inner boundary (piston), diamonds mark instances of time for which snapshots of the atmospheric structure were stored by the radiation-hydrodynamics code. Denoted are selected phases for which line profiles are presented in Fig.\ref{f:co16mueprofiles-WHyavgl}. \textit{Other panels:} Atmospheric structures of the initial hydrostatic model (thick black line) and selected phases $\phi_{\rm bol}$ of the dynamic calculation (colour-coded in the same way as the phase labels of panel a). Plotted are gas temperature (b), gas density (c), and gas velocities (d). Note that this is a dust-free model and the degree of condensation of carbon into dust is $f_{\rm c}$\,=\,0 for every depth point at each instance of time.}
\label{f:structureSRV}
\end{figure*}

\begin{figure*}
\centering
\includegraphics[width=17cm]{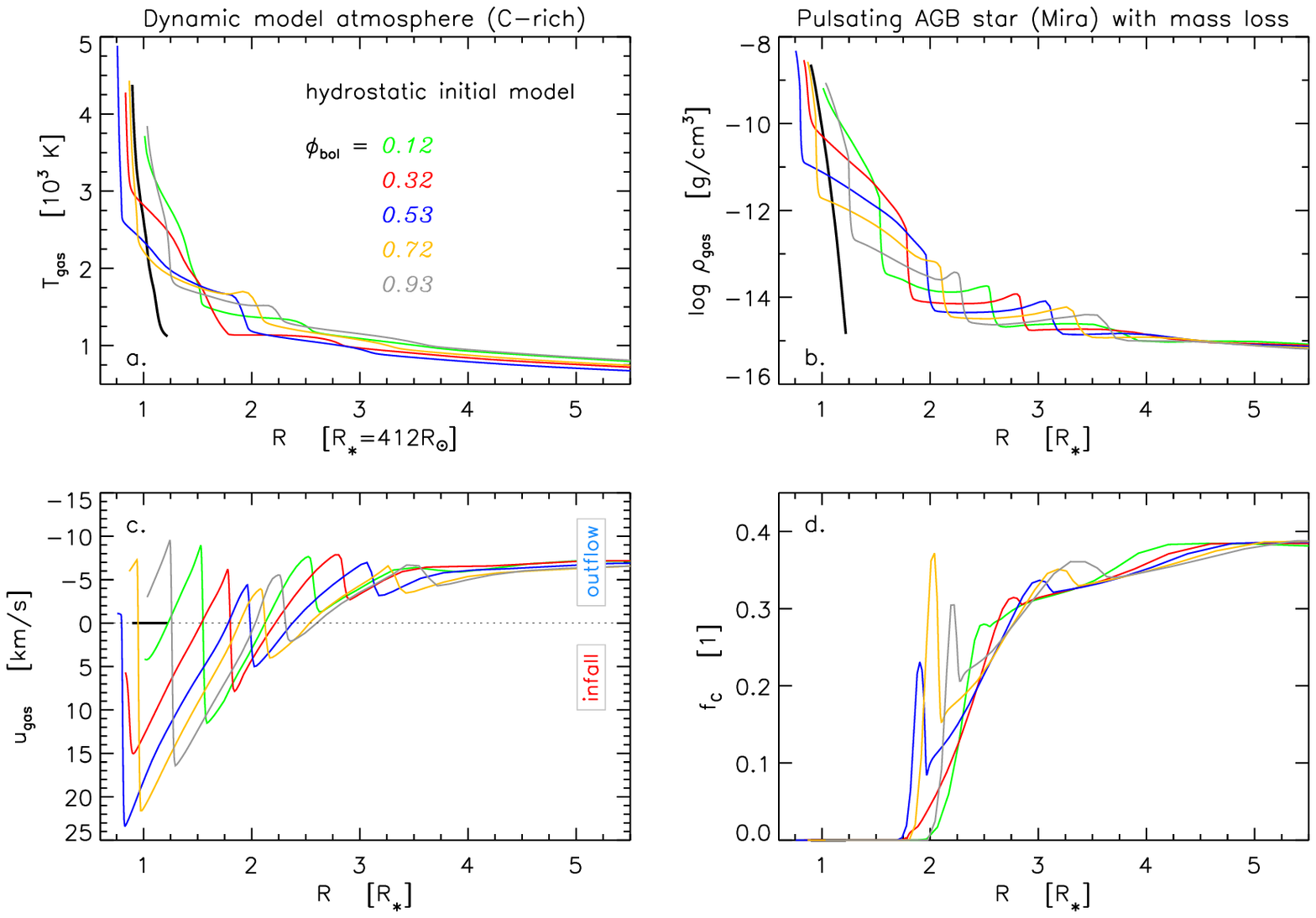}
\caption{Characteristic properties of model~M. Radial structures of the initial hydrostatic model (thick black) and selected phases of the dynamic model atmosphere during one pulsation cycle ($\phi_{\rm bol}$=\textit{0.0--1.0}). Plotted are gas temperatures (a), gas densities (b), gas velocities (c), and degrees of condensation of carbon into dust (d). Panel b demonstrates the larger extension compared to the hydrostatic initial model and the local density variations.}
\label{f:structure}
\end{figure*}

Following the previous remarks, we concentrated in this study on model atmospheres for C-type LPVs for our line profile modelling, because these models contain a more consistent prescription of dust formation.\footnote{First M-type models were presented by Jeong et al. (\cite{JeWLS03}). However, the development of a wind driven by radiation pressure on dust grains of O-rich species in these models possibly results from an overestimation of dust opacities in the course of grey radiative transfer as applied by these specific models (Woitke \cite{Woitk06b}, H\"ofner \cite{Hoefn07}).} However, the velocity effects on line profiles are of general relevance, and a comparison with observational results of stars with other spectral types (e.g. $\chi$\,Cyg) should be justified. This is especially the case for CO lines because of the characteristic properties of this molecule (e.g. Sect.\,2.2 in Paper\,I). Table\,\ref{t:dmaparameters} lists the parameters and the resulting wind properties of the dynamic models used in this work.

The starting point for the calculation is a hydrostatic initial model, which is very similar to classical model atmospheres (Fig.\,1 in DMA3), as for example those calculated with the MARCS-Code (Gustafsson et al. \cite{GEEJN08}). These initial models are characterised by a set of parameters as listed in the first part of  Table\,\ref{t:dmaparameters}. They are dust-free and rather compact in comparison with fully developped dynamical structures (e.g. Fig.\,\ref{f:structure}). The effects of pulsation of the stellar interior are then simulated by a variable inner boundary 
\begin{equation}\label{e:pistonmovement}
R_{\rm in}(t)=R_{\rm in}(0)+\frac{\Delta u_{\rm p}P}{2\pi}\sin\left(\frac{2\pi}{P}t\right) .
\end{equation}
This so-called \mbox{\textit{piston}} moves sinusoidally with a period $P$ and a velocity amplitude $\Delta u_{\rm p}$. Assuming a constant radiative flux at the inner boundary, the luminosity $L_{\rm in}(t)$ varies like $\propto$\,$R_{\rm in}^2(t)$ for the models presented in DMA3. The assumptions for the inner boundary were slightly adapted for the latest generation of models as discussed in DMA4: an additional free parameter $f_{\rm L}$ was introduced so that the luminosity at the inner boundary varies like
\begin{equation}\label{e:flfunction}
\frac{L_{\rm in}(t)-L_{\rm 0}}{L_{\rm 0}} = f_{\rm L}\left(\frac{R_{\rm in}^2(t)-R_{\rm 0}^2}{R_{\rm 0}^2}\right) ,
\end{equation}
where $L_{\rm 0}$ and $R_{\rm 0}$ are the values at the inner boundary of the hydrostatic initial model ($L_{\rm 0}$=$L_\star$, by definition, and $f_{\rm L}$=1 corresponds to the previously used case of constant flux at the inner boundary).\footnote{Note that the average $L$ over a period is slightly higher in a dynamical model than in the corresponding hydrostatic initial model due to the quadratic dependence of $L_{\rm in}(t)$ on $R_{\rm in}(t)$. The effect is small for typical pulsation amplitudes,though ($\approx$\,0.1\% for model~W, $<$\,2\% for model~M). Note also that $R_{\rm 0}$=$R_{\rm in}(0)$ is not equal to $R_\star$ (calculated as described in Tab.\,\ref{t:dmaparameters}), but somewhat smaller, as the inner boundary is located below the photopshere (e.g. for model~S: $R_{\rm 0}$=2.832157$\cdot$10$^{\rm 13}$cm, $R_\star$=3.42803$\cdot$10$^{\rm 13}$cm).}
The luminosity variation amplitude of the model can be adjusted thereby independent of the mechanical energy input by the piston. This allows us to tune the bolometric amplitudes $\Delta m_{\rm bol}$ to resemble more closely the values derived from observational studies. Figure\,\ref{f:structureSRV}a shows the variability of the luminosity $L$ of one dynamic model as an example. 

As described in detail in DMA3, the models can be divided into two sub-groups according to their dynamical behaviour:\\

\begin{itemize}
\item[$\triangleright$]
pulsating model atmospheres where no dust forms
\item[$\triangleright$]
models developing pulsation-enhanced dust-driven winds.
\end{itemize}

\textit{Model~W} (directly taken from DMA3 where it is denoted by l70t28c14u2) represents a typical example for the first type of dynamical models. Figure\,\ref{f:massenschalenW} shows the temporal evolution of mass layers in different atmospheric depths for such a dust-free, pulsating atmosphere. A mild pulsation is simulated by the small amplitude of the piston in this case. However, the matter is not ejected far enough so that dust formation is made possible by the low temperatures. There is no dust-driven outflow, and pulsation is the sole reason for atmospheric dynamics. The outer boundary follows the movement of the upper atmospheric layers in the case of models without mass loss. A shock wave arises once in every pulsation period, causing the layers to follow more or less ballistic trajectories. The resulting completely periodic behaviour can be recognised in Fig.\,\ref{f:massenschalenW}. The atmospheric structure for different phases throughout a pulsational period is shown in Fig.\,\ref{f:structureSRV}. Varying around the hydrostatic configuration, the dynamic model does not resemble it at any point. 

The atmospheres of LPVs become qualitatively different with the occurence of dust and the development of a stellar wind. The other models listed in Table\,\ref{t:dmaparameters}, namely \textit{models S} and \textit{M}, represent examples for this second type of dynamic model atmospheres. Again, shock waves are triggered by the pulsating stellar interior and propagate outwards. The difference arises because efficient dust formation can take place in the wake of the shock waves (post-shock regions with strongly enhanced densities at low temperatures). Radiation pressure on the newly formed dust grains results in an outflow of the outer atmospheric layers. This behaviour is demonstrated in Fig.\,\ref{f:massenschalenM}, which shows the characteristic pattern of the moving mass layers with model~M as an example. While the models without mass loss stay rather compact compared to the hydrostatic initial model (Fig.\,\ref{f:structureSRV}), the models that develop a wind are inflated by it and become much more extended than the corresponding initial model. Numerically, a transmitting outer boundary -- allowing outflow; fixed at 20--30\,$R_{\star}$ -- is used for the latter type of dynamic model atmospheres. Spatial structures of model~M are shown in Fig.\,\ref{f:structure}, demonstrating the strong influence of dust formation on the atmospheric extension. The models S and M exhibit quite a moderate dust formation process, which leads to a smooth transistion of the velocity field from the pulsating inner layers to the steady outflow of the outer layers (see the velocity structure plots in Fig.\,9 of Paper\,I and Fig.\,\ref{f:structure} in this work). In contrast, the model applied in Sect.\,6.1 of Paper\,II is a representative of a group of models with a more extreme dust formation. Not every emerging shock wave leads to the formation of dust grains (cf. Fig.\,2 in DMA3) and the velocity field in the dust-forming region may look very different for similar phases of different pulsation periods. Pronounced dust shells arise from time to time and propagate outwards (see the structure plot in Fig.\,10 of Paper\,II). 

We refer to H{\"o}fner et al. (\cite{HoGAJ03}) for more details about the numerical methods and the atmospheric models.

\subsection{The specific models used}\label{s:DMAsusedspecific}

Model~S was already used extensively in the previous Papers\,I+II to study line profile variations. Its parameters (listed in Table.\,\ref{t:dmaparameters}) were chosen to resemble the Mira S\,Cep, as this is the only \mbox{C-type} Mira with an extensive time series of high-resolution spectroscopy and derived RVs (Fig.\,\ref{f:rvs-chicyg-scep}). Table\,1 of Paper\,II lists properties for this object as compiled from the literature for comparison. Model~S should not be taken as a specific fit for S\,Cep, though. A discussion of the difficulties when relating dynamic model atmospheres to certain objects can be found in Sect.\,3 of Paper\,II (or Sect.\,2.1.4 in N05). However, there is evidence that the model reproduces the outer layers of this star reasonably well. There are some properties listed in the mentioned tables which can be compared directly and agree to some extent (e.g. $L$, $P$, $\dot M$, $u_{\rm exp}$). In addition, low-resolution synthetic spectra in the visual and IR computed on the basis of model~S resemble observed spectra (ISO, KAO) of S\,Cep fairly well as it was shown by Gautschy-Loidl et al. (\cite{GaHJH04}; their Sect.\,5.1).

Model~M was the (so far) last model of a small parameter study we carried out subsequent to Papers\,I+II. The intention was not to find a model fitting a certain target of observations better, but to change the model parameters in order to reproduce one particular observational aspect: namely the velocity variations in the inner, dust-free photosphere resulting in a rather uniform RV curve for CO $\Delta v$\,=\,3 lines in spectra of Miras as discussed in Sect.\,\ref{s:largerampl} and shown in Fig.\,\ref{f:rv-CO-dv3-allMiras}. The comparison of the global velocity field of this model~M with observational results in Sect.\,\ref{s:globalvelfield} is still done on a qualitative basis, though. Confronting RV measurements (Fig.\,\ref{f:rvs-chicyg-scep}) with the corresponding synthetic values (Fig.\,\ref{f:rvsmodelM}) provides some general information of how realistic the velocity structures of the model are. However, a direct and quantitative comparison of stellar and model parameters is not feasible at the moment due to limitations on the observational side (very small number of stars observed extensively, often only rough estimates for properties of targets) as well as on the modelling side  (only C-rich models, laborious process to get a RV diagramm as Fig.\,\ref{f:rvsmodelM}) as discussed in Sect.\,\ref{s:globalvelfield}.

While investigating synthetic CO $\Delta v$\,=\,3 line profile variations based on a few dynamical models available at that time, we found model~W reproducing the observed behaviour of SRVs (Sect.\,\ref{s:COdv3SRVsobs}) with similarities to W\,Hya (increased $\Delta RV$). In Sect.\,\ref{s:COdv3modelW} we will make a comparison of observational and modelling results only for this very selected aspect. Relating model~W (C-rich chemistry) and the M-type LPV W\,Hya to constrain the parameters of this star is even less possible than for the before mentioned case of model~M, and was not intended.

\subsection{Calculating synthetic line profiles}\label{s:synthesis}

For the spectral synthesis we followed the numerical approach as described in detail in N05 (Sect.\,2.2) and also in Papers\,I+II. The modelling procedure of dynamic model atmospheres (see DMA3) yields some immediate results, like mass loss rates $\dot M$, terminal velocities of the winds $v_\infty$, or degrees of dust condensation in the outflows $f_c$. In addition, it provides snapshots of the time-dependent atmospheric structure ($\rho$, $T$, $p$, $u$, etc.) at several instances of time (e.g. Figs.\,\ref{f:structureSRV}+\ref{f:structure}). These represent the starting point for the aspired spectral synthesis, accomplished in a two-step process as described below.

The first step, calculating opacities based on a given atmospheric structure, was accomplished with the COMA code, a description of which can be found in Aringer (\cite{Aring00}), Gautschy-Loidl (\cite{Gauts01}), or N05. Informations on recent updates and the latest version can be found in Gorfer (\cite{Gorfe05}), Lederer \& Aringer (\cite{LedeA09}), Aringer et al. (\cite{AGNML09}). Element abundances for the spectral synthesis were used in consistency with the hydrodynamic models of solar composition. We adopted the values from Anders \& Grevesse~(\cite{AndeG89}), except for C, N and O where we took the data from Grevesse \& Sauval~(\cite{GrevS94}). This agrees with our previous work (e.g. Aringer et al.~\cite{AHWHJ99}, Aringer et al. \cite{AGNML09}) and results in $Z_{\odot}$\,$\approx$\,0.02. Subsequently, the carbon abundance was increased according to the C/O of the models. Abundances and ionisations of various atoms and formed molecules for all layers of the atmospheric model were calculated with equilibrium chemistry routines (for a detailed discussion and an extensive list of references we refer to Lederer \& Aringer \cite{LedeA09}). The depletion of carbon in the gas phase due to consumption by dust grain formation is also taken into account by the COMA code. Examples for the resulting partial pressures can be found in Fig.\,\ref{f:pp}. Subsequently, opacities for every radial depth point and the chosen wavelength grid were computed. Several opacity sources were considered, the most important one being the molecules for which the line profiles are to be studied. Their contribution to the opacities were computed by using line lists and under the following assumptions: (i) conditions of LTE, (ii) a microturbulence velocity of $\xi$=2.5\,km\,s$^{-1}$, and (iii) line shapes described by Doppler profiles. Assuming LTE conditions, level populations can be computed from Boltzmann distributions at the corresponding gas temperature $T$. The opacity $\kappa_{\nu}$ at a given frequency $\nu$ for a certain transition from state $m$ to state $n$ can then be written as
\begin{equation}\label{e:kappaLL} 
\kappa_{\nu}\ =\ \frac{N\pi e^2}{m_{\rm e} c} \ \frac{gf}{Q(T)} \ \textnormal{e}^{-E_0/kT} \left( 1-\textnormal{e}^{-h\nu_0/kT}\right) \varphi(\nu) ,
\end{equation}
with the number density of relevant particles $N$, the charge $e$ and mass $m_e$ of the electron, the speed of light $c$, the Boltzmann constant $k$, and the energy $h\nu_0$ of the respective radiation. The partition function $Q(T)$ is the weighted sum of all possible states. $E_0$ represents the excitation energy of the level $m$ (from ground state) and $gf$ is the product of the statistical weight $g_{(m)}$ of the level times the oscillator strength $f_{(m,n)}$ of the transition. Line lists usually contain frequencies $\nu_0$ (or in practice wavenumbers), excitation energies $E_0$, and $gf$ values together with informations for line identification. In order to reproduce the line shapes in a realistic way, a broadening function $\varphi(\nu)$ for the line profile was introduced with
\begin{equation}
\int^{\infty}_{0}\varphi(\nu) \ \mathrm{d}\nu=1 \ .
\end{equation}
Only the effects of thermal broadening (first term in Eq.\,\ref{e:dopplerwidth}) and the non-thermal contribution of microturbulent velocities (second term in Eq.\,\ref{e:dopplerwidth}) are taken into account by COMA, whereas other effects (e.g. natural and pressure broadening, macroturbulence) are neglected. We refer to Sect.\,2.2.2 of N05 for details. The resulting Doppler profiles can be described by a (Gaussian) broadening function
\begin{equation}
\varphi(\nu)=\frac{1}{\Delta_\nu \sqrt{\pi}} \  \textnormal{e}^{-\left(\frac{\nu-\nu_0}{\Delta_\nu}\right)^2} , 
\end{equation}
with a Doppler width $\Delta_\nu$ given by
\begin{equation}\label{e:dopplerwidth}
\Delta_\nu=\frac{\nu_0}{c} \ \sqrt{\frac{2\mathcal{R}T}{\mu}+\xi^2} ,
\end{equation}
where $\mathcal{R}$ is the gas constant, $\mu$ the molecular weight, and $\xi$ the microturbulent velocity. An updated set of references for the line lists of all molecular species used by the current version of COMA can be found in Lederer \& Aringer (\cite{LedeA09}). In this work we made use of the list of Goorvitch \& Chackerian (\cite{GoorC94}) for CO and the CN line list of J{\o}rgensen \& Larsson (\cite{JorgL90}). We utilised the same molecular lines as in Papers\,I+II, their respective properties are summarised in Table\,\ref{t:chosenlines}. In addition, molecules affecting the spectra pseudo-continuously (mainly C$_2$H$_2$) were considered by an opacity of constant value for the considered spectral range (cf. Sect.\,5.1. in Paper\,I or N05). Furthermore, continuum absorption coefficients are determined by COMA (Lederer \& Aringer \cite{LedeA09}) as well as the opacity due to dust grains of amorphous carbon.  For the latter we used the data of Rouleau \& Martin (\cite{RoulM91}; set AC) and computed the resulting  dust absorption\footnote{Scattering was not taken into account, neither for the dynamic models nor for the spectral synthesis.\rm} under the assumption of the small particle limit of the Mie theory (grain sizes much smaller than relevant wavelengths $\leadsto$ opacities proportional to the total amount of condensed material, but independent of grain size distribution; cf. DMA3, H\"ofner et al. \cite{HoefnD92}).

In the second step of the spectral synthesis, the previously calculated data array of opacities $\kappa_{\nu}$($\lambda$,r) for all depth and wavelengths points was utilised to solve the radiative transfer (RT). As the thickness of the line-forming region in AGB atmospheres is large compared to the stellar radii, it is necessary to treat the RT in spherical geometry. In addition, the complex velocity fields (pattern of outflow and infall as for example shown in Fig.\,\ref{f:structure}) in AGB atmospheres severely affect the line shapes in the resulting spectra (observed or synthetic, e.g. Fig.\,\ref{f:complexlineformation}), and it is essential to include the influence of relative  macroscopic velocities in this step of the spectral synthesis. Thus, a code for solving spherical RT, which takes into account velocity effects, is used to model line profiles and their variations. The RT code used in this work (Windsteig \cite{Winds98}) follows the numerical algorithm described in Yorke (\cite{Yorke88}).

It is necessary to choose spectral resolutions that are high enough to sample individual spectral lines with a sufficient number of wavelength points. This is especially important for synthesising the often quite complex line profiles for stars with pronounced atmospheric dynamics, like those that are the topic of this work. Therefore, all spectra are calculated with an extremely high resolution of $R$=$\lambda/\Delta\lambda$=300\,000 and were then rebinned to $R$=70\,000 for comparison with observed FTS spectra. Furthermore, the synthetic spectra shown below were normalised relative to a computation with only the continuous opacity taken into account ($F$/$F_{\rm cont}$). 

Our aim is to infer information about atmospheric velocity fields from shifts in the wavelength of (components of) spectral lines. For this purpose, RVs were calculated by using the rest wavelength of the respective line and the formula for Doppler shift. For an easy comparison of our modelling results with observations, we adopted the naming convention for velocities of observational studies (positive for material moving away from the observer and negative for matter moving towards the observer). Thus, outflow from the star results in blue-shifted lines and negative RVs, while infalling matter revealed by red-shifted lines leads to positive RVs. In general, the RVs resulting from observations were combined into one composite lightcycle \mbox{($\phi_{\rm v}$=0.0--1.0)} and then plotted repeatedly for better illustration (e.g. Fig.\,\ref{f:rv-CO-dv3-allMiras}). For the modelling we computed spectra and derived RVs throughout one pulsation period \mbox{($\phi_{\rm bol}$=\textit{0.0--1.0})} and replicate the values beyond this interval (e.g. Fig.\,\ref{f:rv-CO-dv3-modelsMS}).

Following the convention of Papers\,I+II, \textit{bolometric phases} $\phi_{\rm bol}$ within the lightcycle in luminosity (Fig.\,\ref{f:structureSRV}a) will be used throughout this work to characterise the modelling results (atmospheric structures, synthetic spectra, derived RVs, etc.) with numbers written in \textit{italics} for a clear distinction from the visual phases $\phi_{\rm v}$, which are usually used to denote observational results (with $\phi_{\rm v}$=0 corresponding to phases of maximum light in the visual). For a discussion of the relation between the two types of phase informations, namely $\phi_{\rm bol}$ and $\phi_{\rm v}$, we refer to Appendix\,\ref{s:phaseshift}.

Throughout this work (e.g. the lower panel of Fig.\,\ref{f:kappatauauswahl} and all similar plots in the following), the radial optical depth $\tau_{\nu}$(r) is computed by radially integrating inwards\footnote{corresponding to the central beam with impact parameter $p$=0 in the framework of spherical RT (e.g. Fig.\,1 of Nordlund \cite{Nordl84})} for a given wavelength point 
\begin{equation}\label{e:optdepth}
\tau_{\nu}(r)=\int^{r}_{r_{\rm max}} [\kappa_{\nu,\rm cont}(r')+\kappa_{\nu,\rm lines}(r')+\kappa_{\nu,\rm dust}(r')]\,\, \rho(r')\, \mathrm{d}r'
\end{equation}
and is then plotted with dotted lines. For the opacities $\kappa_{\nu}(r)$, all relevant sources (continuous, molecular/atomic, dust) are included. If velocity effects are taken into account, optical depths are calculated with the opacities at all depth points Doppler-shifted according to the corresponding gas velocity there and then plotted with solid lines. The optical depth of $\tau_{\nu}$$\approx$1 provides clues on the approximate location of atmospheric layers where radiation of a certain frequency $\nu$ (or the corresponding wavelength $\lambda$) originates.

\begin{table}
\begin{center}
\caption{Different molecular features and the properties of the specific lines
used for the line profile modelling here (and in Papers\,I+II).}
\begin{tabular}{llcc}
\hline\hline
&\multicolumn{3}{c}{Specific line chosen for modelling}\\
\cline{2-4}
\multicolumn{1}{c}{Line type}&designation&$\sigma / wn$ [cm$^{-1}$]&$\lambda$
[$\mu$m]\\
\hline
CN $\Delta v$\,=\,--2  red&1--3 Q$_2$4.5&4871.3400&2.0528\\
CO $\Delta v$\,=\,3&5--2 P30&6033.8967&1.6573\\
{\tiny  \it CO $\Delta v$\,=\,2  high-exc.}&{\tiny \it 2--0 R82}&{\tiny \it 4321.2240}&{\tiny \it 2.3142}\\
\hline
CO $\Delta v$\,=\,2  low-exc.&2--0 R19&4322.0657&2.3137\\
\hline
CO $\Delta v$\,=\,1&1--0 R1&2150.8560&4.6493\\
\hline
\end{tabular}
\label{t:chosenlines}
\end{center}
Notes: The lines were chosen according to the criteria given in Sect.\,5.1 of Paper\,I. The values given here correspond to the rest wavelengths or central frequencies $\nu_0$ as used in Eq.\,(\ref{e:kappaLL}). Throughout this work we will use only a low-excitation first overtone CO line (i.e. CO 2--0 R19) denoted by CO $\Delta v$\,=\,2, while the high-excitation line CO 2--0 R82 is merely listed for completeness as it was used in Papers\,I+II. This distinction is necessary as the two types of lines (at least the ones used by HHR82 and similar studies) behave differently in kinematics studies (see Fig.\,\ref{f:rvs-chicyg-scep}). 
\end{table}

\section{Line formation in the extended atmospheres of evolved red giant stars} 
\label{s:lineformation}

This section is concerned with the fact that different molecular spectral features visible in spectra of AGB stars originate at different atmospheric depths and the examination of this with numerical methods.

Typical main sequence stars exhibit relatively compact photospheres, and the radiation in different wavelengths originates at roughly the same geometrical atmospheric depths with a rather well defined temperature (e.g. Sect.\,4.3.1 in GH04). For example, the relative thickness of the flux-forming region where the spectrum is produced in our Sun amounts to $\Delta R_{\rm phot,\odot}/R_{\odot}$\,$\approx$\,4$\cdot$10$^{-4}$ (cf. N05). The often used parameter effective temperature $T_{\rm eff}$ represents a mean temperature in the layers of the thin photosphere, from where almost all photons can escape. 

In contrast, estimating the atmospheric extensions $\Delta R_{\rm atm,\star}$ in the case of AGB stars is much more difficult. The outer boundaries of these objects are hard to define (as discussed by GH04 in their Sect.\,4.1) due to the shallow density gradients related to the low surface gravities $g_\star$, intensified by the dynamic processes of pulsation and mass loss. GH04 give estimates of \mbox{0.1--0.5} for the parameter $\Delta R_{\rm atm,\star}$/$R_{\star}$. For the dynamic models used here (Sect.\,\ref{s:DMAsused}), one could for example consider the region of dust formation at $\approx$2\,$R_{\star}$, where the onset of the stellar wind takes place (cf. Fig\,\ref{f:structure}) as an outer boundary of the atmosphere. This results in $\Delta R_{\rm atm,\star}$/$R_{\star}$\,$\approx$\,1. Whatever value one adopts, it is clear that the thickness of an AGB atmosphere is of the same order as the stellar radius, which is quite different from normal main sequence stars. With these extremely extended atmospheres, the wavelength-dependent geometrical radius can vary by a few 100\,$R_{\odot}$ (equivalent to a few \textit{AU}) or more. Also the related gas temperatures can cover a wide range from $\approx$5000\,K down to a few hundred Kelvin (N05). 

\begin{figure}
\resizebox{\hsize}{!}{\includegraphics[clip]{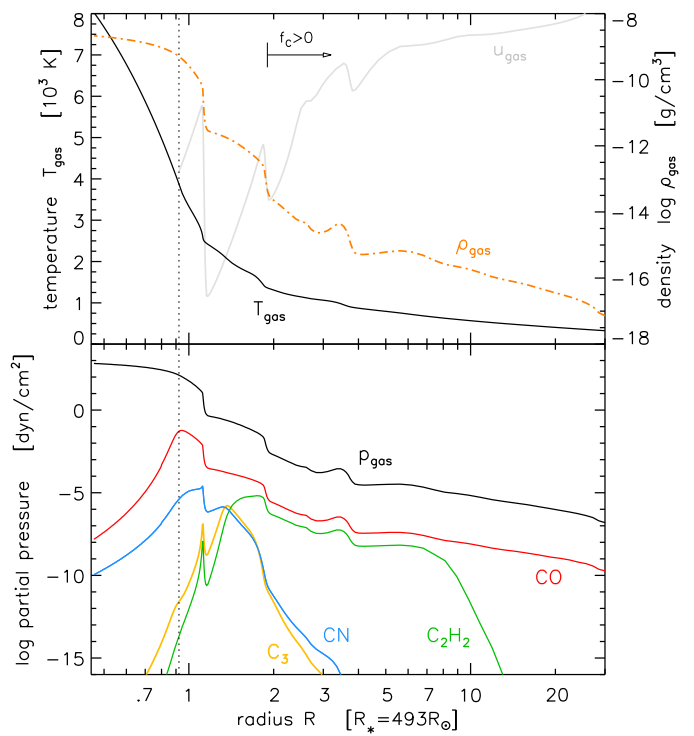}}
\caption{\textit{Upper panel:} Atmospheric structure of model~S for the phase $\phi_{\rm bol}$=\textit{0.0}. Gas velocities $u_{\rm gas}$ are overplotted for illustration purposes (shock fronts). Compare Figs.\,1 and 9 in Paper\,I to get an idea of the temporal variations and also for the scale of the gas velocities. The dotted line marks the border between the actual model and the extension towards the interior for the sake of optical depth (see text). \textit{Lower panel:} The total gas pressure together with the corresponding partial pressures of selected molecules for this phase if chemical equilibrium is assumed. Shown is only a subset of all molecular species considered by COMA, which are CO, CN, C$_2$, CH, C$_3$, C$_2$H$_2$, and HCN.}
\label{f:pp}
\end{figure}

In general, molecular lines do not originate in a well-defined and narrow region, but are formed over a more or less wide range in radius. Nevertheless, there have been attempts to constrain the approximative line forming region in radius and temperature for different molecular features with similar dynamical model atmospheres as used here. The results -- which should be regarded as rough estimates rather than definite borders, though -- are compiled e.g. in Table\,1.2 of N05. There are only a few molecular bands (e.g. CN or C$_2$), which are solely formed in the deep, warm layers of the photosphere. Most of the strong features (e.g. H$_2$O, HCN, C$_2$H$_2$) originate in the cool, upper layers of the atmosphere because of the small dissociation energies of the (polyatomic) molecules in combination with the high $gf$ values. It is interesting that there are examples for bands formed in a relatively narrow region, as e.g. the ones of C$_3$, which is due to the interplay between molecular formation and dissociation probabilities (Gautschy-Loidl \cite{Gauts01}, Fig.\,\ref{f:pp}). The special role of CO and the corresponding features will be discussed in detail below. To sum this up, the emerging spectrum (across the whole spectral range from the visual to the IR) of an AGB star is formed over a large radial range with diverse physical conditions, which is quite different from the scenario for the Sun as sketched above.

Moreover, not only spectral features of different molecules originate at varying radial atmospheric depths, but also different lines of the same molecular species can show this effect. This is especially pronounced in the case of CO lines and will be investigated below on the basis of one (arbitrarily chosen) phase of model~S. The upper panel of Fig.\,\ref{f:pp} shows the corresponding atmospheric structure\footnote{Note: The dynamical models reach into geometrical depths with temperatures of about 4--5000\,K (Fig.\,\ref{f:structure}). Modelling even deeper zones would (besides numerical difficulties) mean to include convection and the driving mechanism of pulsation, which has not been done so far. The resulting overall continuous optical depth is too low ($<$1) at certain wavelengths, but this is not the case for the wavelengths of strong absorption features, as illustrated in Nowotny et al. (\cite{NAGHL03}). Thus, a linear extrapolation in $T$-$p$ to deeper layers ($T_{\rm gas}$$\approx$10$^4$\,K, $\tau$$\approx$100) had to be applied for the spectral synthesis presented here. Velocities were considered to be zero in the extension. This assumption is not problematic since these layers are only relevant for continuum optical depths and have no influence on the line profiles as molecular lines are formed further out.} at the phase $\phi_{\rm bol}$=\textit{0.0}.

There are a few factors that are decisive for the line intensities and for the locations ($r$, $T_{\rm gas}$) of the line forming regions of molecular features in AGB atmospheres: 
\begin{enumerate}
\item[i.]
the atmospheric structure itself ($r$-$T$-$p$)
\item[ii.]
the relative abundance (i.e. partial pressure) of the considered molecular species at certain depth points within the atmosphere (formation/depletion)
\item[iii.]
the properties of the respective line (excitation energy $E_{\rm exc}$, strength of the transition $gf$)
\item[iv.]
possibly other opacity sources (local continuous opacity, other molecular lines, dust opacity, pseudo-continuous molecular contributions)
\item[v.]
influence due to the relative macroscopic velocity fields.
\end{enumerate}

Several studies demonstrated in the past that the atmospheric structure ($\rightarrow$\,i.) has substantial impact on spectral features. Compared with hydrostatic model atmospheres, dynamic models have enhanced densities in cool upper layers (Fig.\,\ref{f:structure}). This results in a change of molecular abundance and decreased or increased intensities of molecular features (as outlined in Sects.\,4.7.5 and 4.8.3.2 of GH04). Examples for this effect would be the SiO bands at 4$\mu$m (Fig.\,5 of Aringer \cite{AHWHJ99}), the H$_2$O bands around 3$\mu$m (Fig.\,9 in DMA3), or the combined feature of C$_2$H$_2$ and CN at $\approx$14$\mu$m (Sect.\,4 of DMA4).

The lower panel of Fig.\,\ref{f:pp} demonstrates the differences in relative abundance ($\rightarrow$\,ii.) for various molecules throughout the whole atmosphere. Plotted are the partial pressures for different molecular species resulting from the evaluation under the assumption of chemical equilibrium by the COMA code. CN is a typical representative for species which form in deep photospheric layers below $\approx$1.5\,$R_{\star}$. Values of 2400--5000\,K for the gas temperatures in the line forming regions were assigned to this species by Gautschy-Loidl (\cite{Gauts01}). In contrast, C$_2$H$_2$ shows a quite different behaviour. As they are very sensitive to temperature (dissociation), these molecules can be found from $\approx$1.2\,$R_{\star}$ on outwards, Gautschy-Loidl lists temperatures of $<$\,2400\,K. C$_3$ represents a fairly extreme example. Its formation requires at the same time high densities for high enough collision probabilities with low temperatures in order not to dissociate. This is met only in a rather narrow region around temperatures of 2000\,K. The density variations due to the propagating shock wave (for this phase at a radius slightly larger than 1\,$R_{\star}$) also are of importance, as can be seen in the partial pressures. From $\approx$0.8\,$R_{\star}$ inwards the abundances of all species decrease as molecules are dissociated for temperatures higher than 5000\,K.

\begin{figure}
\resizebox{\hsize}{!}{\includegraphics[clip]{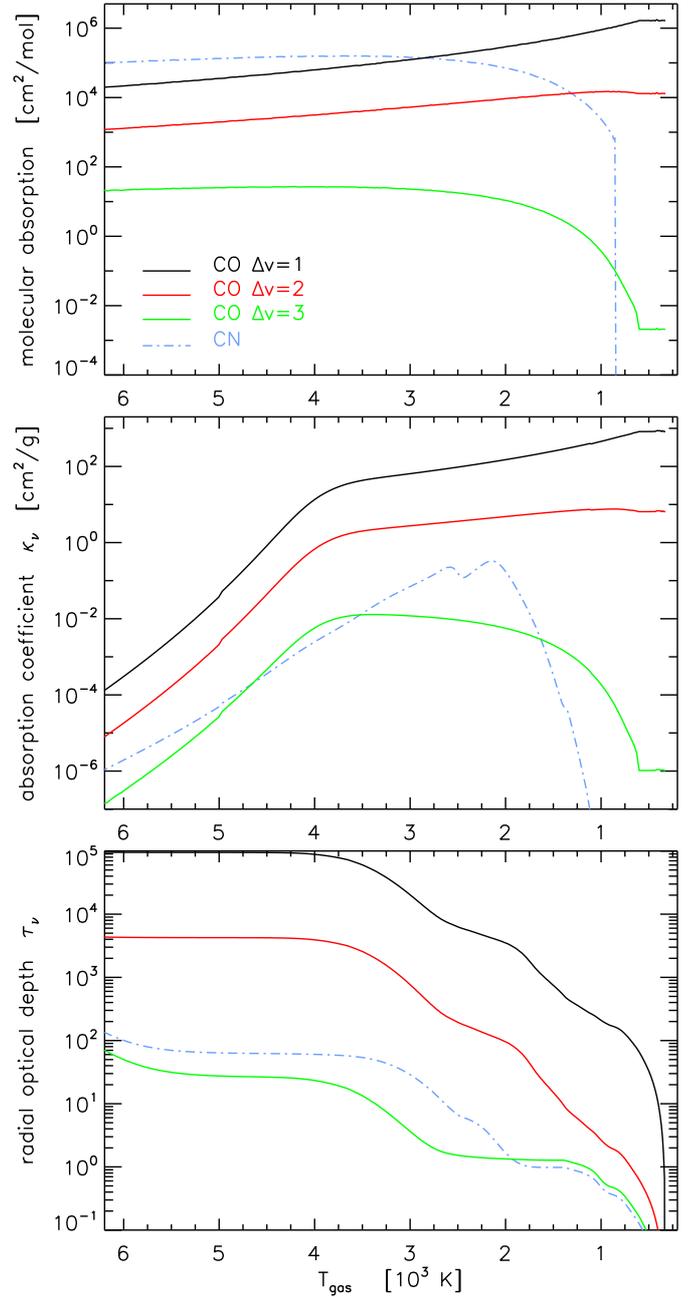}}
\caption{Radial distribution of absorption per mol of the molecular material (upper panel), absorption coefficient (centre panel) and the optical depths (see text for detailed discussion) for the molecular lines chosen for the line profile modelling as listed in Table\,\ref{t:chosenlines}, calculated on the basis of the atmospheric structure shown in the upper panel of Fig.\,\ref{f:pp}. For the absorptions (upper + middle panel) only the contribution of the respective lines are plotted, while all relevant opacity sources (cf. Eq.\,\ref{e:optdepth}) are taken into account for the optical depths (lower panel). Note that the effects of gas velocities (Doppler-shifts) are not taken into account for the computation of the optical depths.}
\label{f:kappatauauswahl}
\end{figure}

\begin{figure}
\resizebox{\hsize}{!}{\includegraphics[clip]{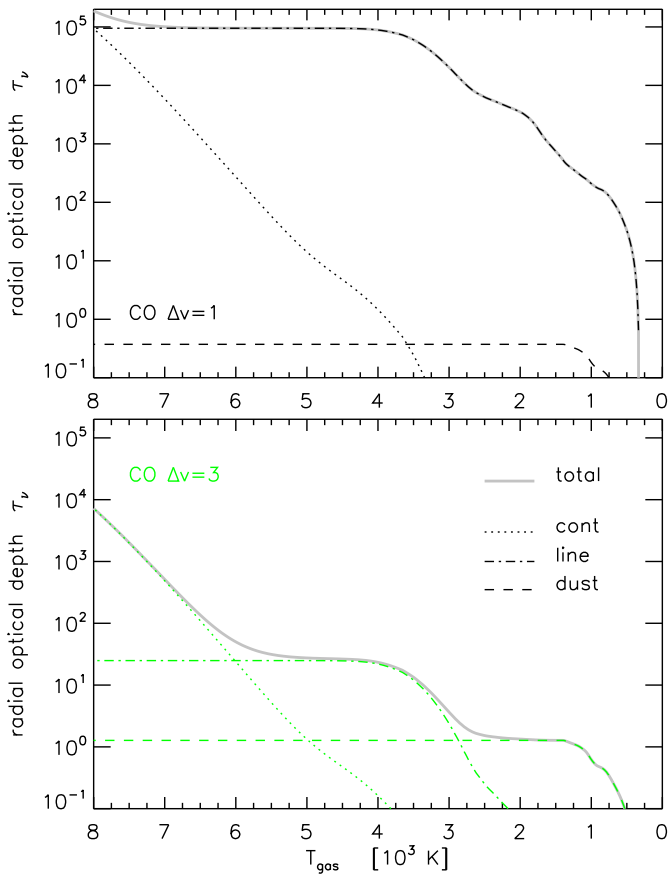}}
\caption{Same as bottom panel of Fig.\,\ref{f:kappatauauswahl}, the individual contributions of different opacity sources (continuous, molecular line, dust) to the radial optical depth (cf. Eq.\,\ref{e:optdepth}) for two of the CO lines used in this study (Tab.\,\ref{t:chosenlines}) are shown in addition.}
\label{f:tauaufgedroeselt}
\end{figure}

\begin{figure}
\resizebox{\hsize}{!}{\includegraphics[clip]{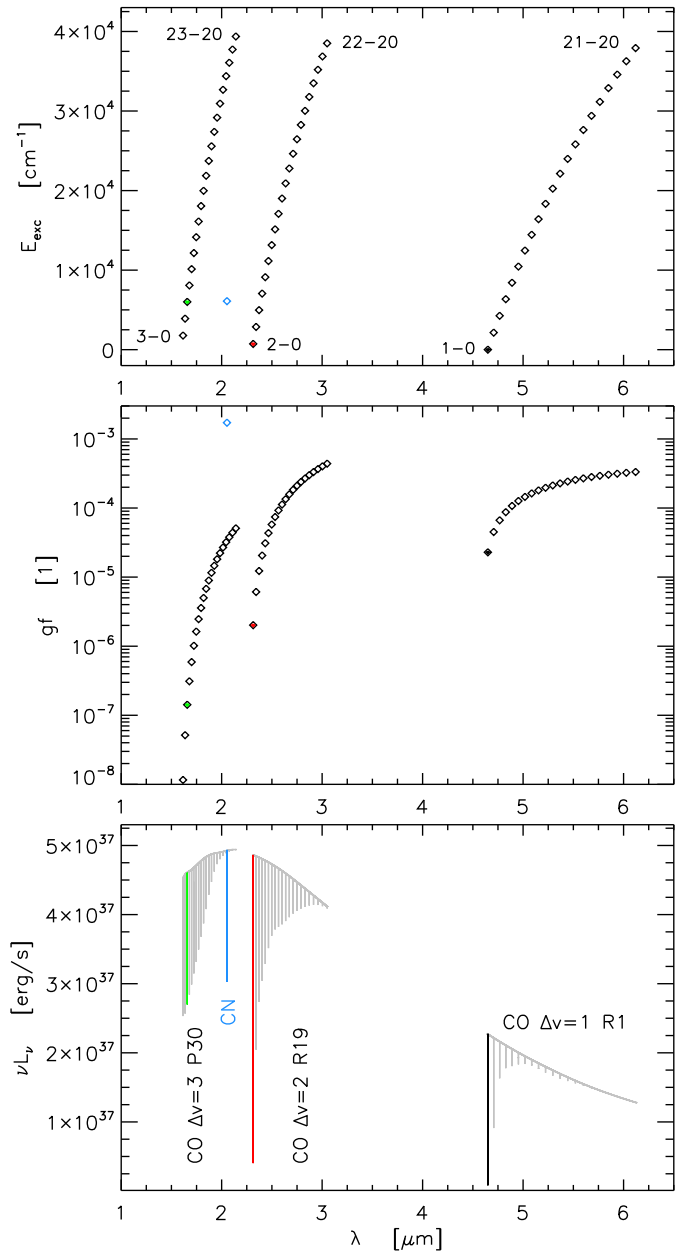}}
\caption{Characteristic properties excitation energy and $gf$ value of several lines of the CO $\Delta v$\,=\,3\,/\,2\,/\,1 bands are shown in the \textit{upper} and \textit{middle panel}. Plotted in each case are lines of the same rotational transition but from different vibrational bands (meaning for example \mbox{1--0 R1}, \mbox{2--1 R1,} 3--2 R1, ... , 21--20 R1). The strong variations in $E_{\rm exc}$ and $gf$ from one vibrational band to another influence the results shown in Fig.\,\ref{f:kappatauauswahl} and lead to distinct different line intensities. This is illustrated in the \textit{lower panel} with a spectrum (frequency times specific luminosity vs. wavelength) based on the atmospheric model of Fig.\,\ref{f:pp} where only the respective lines are taken into account. Marked with the same colour-code as Fig.\,\ref{f:kappatauauswahl} are the lines actually used for the line profile modelling (as listed in Table\,\ref{t:chosenlines}).}
\label{f:linespecifica}
\end{figure}

\begin{figure}
\resizebox{\hsize}{!}{\includegraphics[clip]{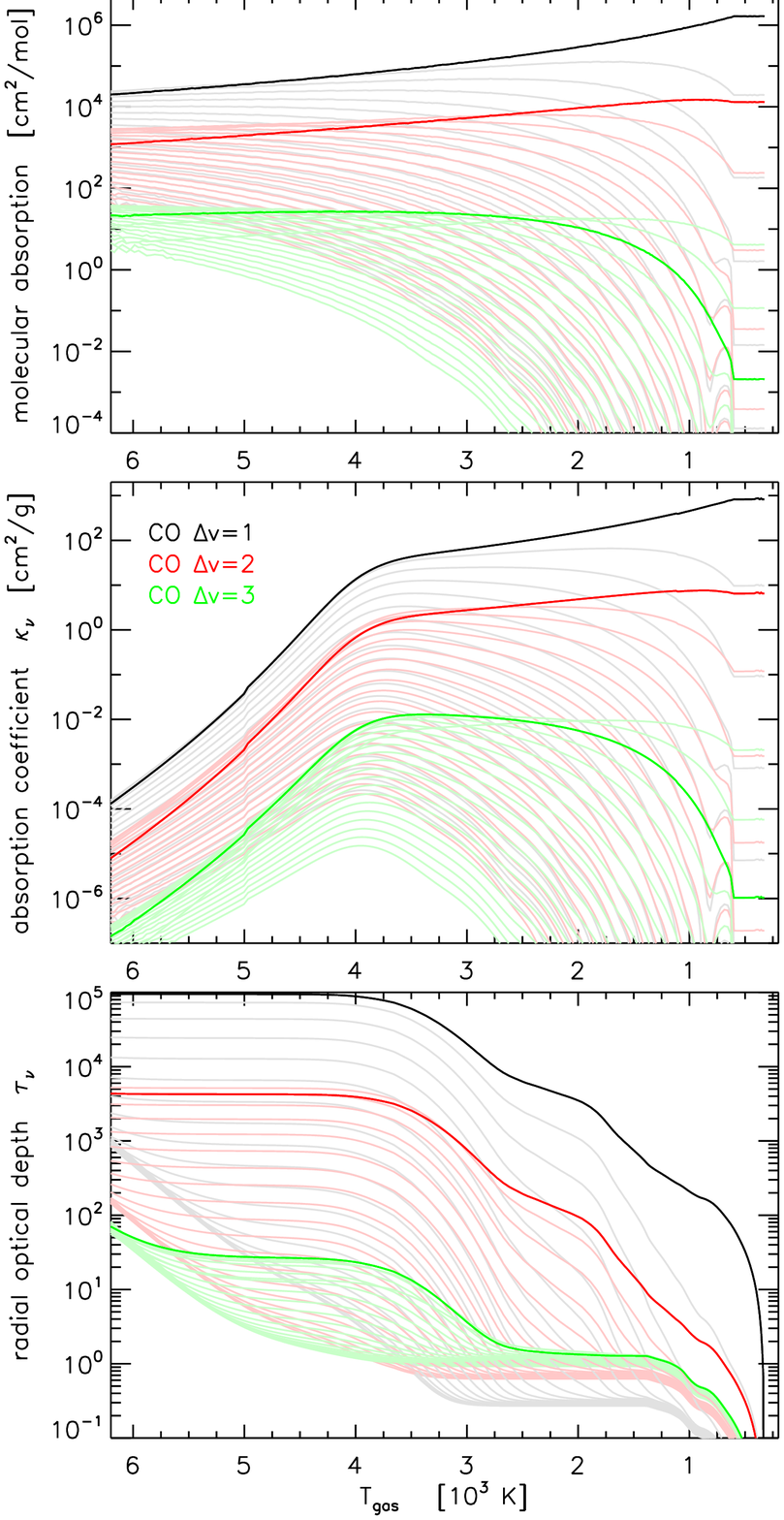}}
\caption{Same as bottom panel of Fig.\,\ref{f:kappatauauswahl} but for the large number of CO $\Delta v$\,=\,3\,/\,2\,/\,1 lines as shown in Fig.\,\ref{f:linespecifica}. Plotted with thick linestyle are the CO lines that were actually used for the line profile modelling (Fig\,\ref{f:kappatauauswahl}). }
\label{f:kappatauvielzahl}
\end{figure}

Figure\,\ref{f:pp} illustrates also that CO is an outstanding molecular species concerning its relative abundance. Due to its high dissociation energy (11.1\,eV), CO is stable and the most abundant species (of those shown; not included in this figure is e.g. the by far most abundant molecular species H$_{\rm 2}$) at all depth points of the atmospheric model. For some molecules -- like CN or even more C$_3$ -- the limited range with significant large partial pressures sets constraints on the line forming region. However, CO is present and abundant across the whole atmosphere, and the properties of individual spectral lines ($\rightarrow$\,iii.), namely the excitation energy $E_{\rm exc}$ and the strength of the transition $gf$, become important. The excitation energy $E_0$ of the lower level of a certain transition $m$\,$\rightarrow$\,$n$ determines at which temperatures (i.e. atmospheric depths) the levels are populated so that an absorption can occur at all. In addition, the strength of the transition \mbox{($gf$-value)} determines the intensity of the line and thus also the atmospheric depth were $\tau_{\nu}$$\approx$1 is reached. Compare Sect.\,\ref{s:synthesis} and Eq.\,(\ref{e:kappaLL}) in this context. Also the relative abundance is present in this equation by the quantity $N$.

To study these effects for the lines chosen for our line profile modelling (cf. Table\,\ref{t:chosenlines}, Fig.\,\ref{f:linespecifica}), we used the atmospheric structure shown in Fig.\,\ref{f:pp} and calculated different quantities. Note that velocity effects were neglected for all subsequent computations of this subsection. The results are plotted in Fig.\,\ref{f:kappatauauswahl}. The upper panel of this plot shows the molecular absorption (absorption per mol of the material of the corresponding molecular species), illustrating how one individual molecule of a given species would absorb radiation of the wavelength of the corresponding line at all atmospheric depth points. Level populations are computed by Boltzmann distributions for the respective temperature (LTE). The increasing $gf$ values from second overtone ($\Delta v$\,=\,3) to first overtone ($\Delta v$\,=\,2) and to fundamental ($\Delta v$\,=\,1) CO lines (cf. Fig.\,\ref{f:linespecifica}) are reflected in the increasing molecular absorption. The middle panel of Fig.\,\ref{f:kappatauauswahl} shows the actual absorption coefficient (absorption per gram of the whole stellar material) due to the respective molecular line, where the relative abundance of the species comes in. This can be recognised by the decrease for all $\kappa_{\nu}$ values with higher temperatures or the steep decrease for the CN line below 2000\,K where the partial pressure of CN severely drops (Fig.\,\ref{f:pp}). Plotted in the middle panel is the opacity due to the CO or CN lines only. In combination with all other contributions (continuous, other molecular species, dust; $\rightarrow$\,iv.), these opacities serve as input for the RT or can be used to calculate radial optical depths as shown in the lower panel of Fig.\,\ref{f:kappatauauswahl}. The quantities $\kappa_{\nu}$ and $\tau_{\nu}$ for the central/rest wavelength of the respective lines (cf. line profiles in Fig.\,\ref{f:complexlineformation} plotted grey) are shown here, no velocity effects are taken into account. It can clearly be seen that $\tau_{\nu}$$\approx$1 is reached in a wide temperature range between $\approx$300\,K and $\approx$3000\,K, corresponding to \mbox{$\Delta R$$\approx$10--20\,$R_{\star}$$\approx$23--46\,\textit{AU}} in this case. While for the CN line the limited region of high enough partial pressure \textit{and} the line parameters are crucial for the line formation, only the latter is relevant for CO lines. Different rotation-vibration band systems of CO originate in quite separated regions within the stellar atmosphere. Therefore and because of other important effects (cf. Sect.\,2.2 of Paper\,I), the spectral lines of CO play a major role in the research on AGB atmospheres, especially for kinematic studies. 

The contributions of different opacity sources to the resulting radial optical depths are illustrated in Fig.\,\ref{f:tauaufgedroeselt} for the model atmosphere used (Fig.\,\ref{f:pp}). The upper panel of this plot shows that the strong fundamental CO line is the dominant source at its central/rest wavelength over almost the whole extended atmosphere. Only in the hot layers below the photosphere the continuous opacity takes over, while the dust plays a neglectable role in general. The panel below shows the same quantities for the chosen second overtone CO line. Because of its properties (lower $gf$ value and higher $E_{\rm exc}$ $\rightarrow$ lower line intensity compared to the $\Delta v$\,=\,1 line; cf. the spectrum in Fig.\,\ref{f:linespecifica}), this molecular line provides the main contribution to the total optical depth only in the layers of $\approx$\,2--4000\,K. From $\approx$\,5000\,K inwards the continuous opacity represents the major source again, although to a somewhat lower extent (minimum of absorption due to the H$^-$ ion around 1.6\,$\mu$m; e.g. GH04) compared to the former case ($\approx$\,4.65\,$\mu$m). On the other hand, the absorption due to grains of amorphous carbon dust is stronger in the wavelength region of CO $\Delta v$\,=\,3 lines (cf. Fig.\,1 of Andersen et al. \cite{AndLH99}) as reflected in the dust optical depths. Still, the absorption features are visible in the synthetic spectra because of the moderate mass loss of the used model~S (Tab.\,\ref{t:dmaparameters}). This changes for higher mass loss rates, the intensities of molecular features decrease as the corresponding line forming regions are hidden by optically thick dust shells.

Figure\,\ref{f:linespecifica} illustrates that the same type of CO lines (fundamental, first or second overtone), but from different vibrational bands can have very different values of $E_{\rm exc}$ and $gf$. Not only leading to variations in line intensities (lower panel), it also has a strong impact on the absorption coefficients and optical depths as discussed above. This is demonstrated in Fig.\,\ref{f:kappatauvielzahl}, where the radial optical depths are plotted for the variety of lines from different vibrational bands used for Fig.\,\ref{f:linespecifica}. Going to higher vibrational quantum numbers leads to weaker lines and line formation regions located at higher temperatures ($T_{\rm gas}$), i.e. further inside the atmosphere. One may suppose that these differences should also be reflected in the dynamic effects (line profile variations, derived RVs) if other lines than those used for our line profile modelling (Table\,\ref{t:chosenlines}; chosen to be similar to observational studies) are considered.

The remaining factor ($\rightarrow$\,v.) concerning the line formation process, namely the important point of the influence of velocity fields, will be the main issue in the next sections. Velocity effects are most relevant for line profiles in spectra of evolved red giants.

\section{The influence of velocity fields} \label{s:velocityeffects}

The previous section covered the line formation process within the extended AGB atmospheres with emphasis on the aspects of atmospheric structures, molecular abundances, and properties of molecular lines. Now, we will introduce another important item relevant for the line formation in the atmospheres of long-period variables. Representing \textit{the} major topic of this work, atmospheric velocity fields with relative macroscopic motions of the order of 10\,km\,s$^{-1}$ are decisive for the resulting profiles of individual spectral lines. This is illustrated below by numerical computations for selected examples of increasing complexity. Compare also Mihalas (\cite{Mihal78}; Sect.\,14) in this context for a discussion of the theoretical background.

\subsection{The simple case}  \label{s:factorp}

\begin{figure}
\resizebox{\hsize}{!}{\includegraphics[clip]{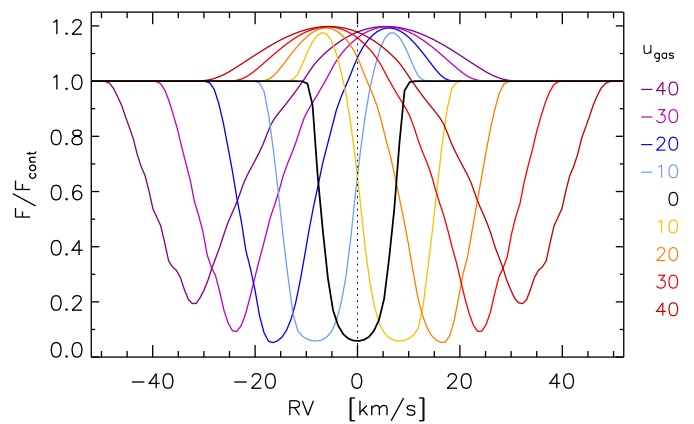}}
\caption{Synthetic CO $\Delta v$\,=\,2 line profiles based on the hydrostatic initial atmospheric structure of the dynamic model~W (Fig.\,\ref{f:structureSRV}). Artificial gas velocities constant for each atmospheric depth point were assumed, these are listed in the legend on the right in units of [km\,s$^{-1}$] and are colour-coded in the same way as the spectra. The velocity field has a strong influence on the resulting line profiles (shapes, intensities, shifts, emission components). Outflow results in blue-shifted absorption profiles, while infall leads to red-shifted ones. Values for the conversion factor \textit{p}, as derived from the Doppler-shifted line profiles shown here, are listed in Table\,\ref{t:factorp}.}
\label{f:dopplershift}
\end{figure}

The first example is shown in Fig.\,\ref{f:dopplershift}, representing the most simple case conceivable for illustration purposes. For this, we used the hydrostatic initial atmospheric structure of model~W (cf. Table\,\ref{t:dmaparameters}). Based on this atmospheric structure, synthetic line profiles for a selected CO $\Delta v$\,=\,2 line (Table\,\ref{t:chosenlines}) were calculated under the assumption of uniform velocity fields, i.e. spherically symmetric outflow or infall with one constant radial velocity for every atmospheric depth point. This leads to profiles shifted to the blue or red, according to the assumed outflow or infall, respectively. However, due to projection effects the gas velocities do not convert directly into observed Doppler-shifts and the measured RVs are always smaller. The shapes of the absorption features change obviously, they become broader and more cone-like but less deep. This effect is increasing with higher velocities. In addition, a weak emission is appearing, Doppler-shifted in the opposite way than the absorption component. This model represents an extended red giant star with outer layers that are not optically thick. Therefore also the surrounding material can significantly contribute to the spectrum, leading to this typical profile of P\,Cygni-type when velocity effects are accounted for.

A correction factor \textit{p} is needed to connect RVs of Doppler-shifted spectral features with actual gas velocities in the corresponding line forming region by $u_{\rm gas}$\,=\,$p$\,$\cdot$\,$RV$. Several studies (e.g. Willson et al. \cite{WilWP82}, Scholz \& Wood \cite{SchoW00}) investigated in the past the size of $p$, as outlined in Sect.\,4.6 of Paper\,II. We also derived values for \textit{p} for the line profiles shown in Fig.\,\ref{f:dopplershift}. For this purpose, the assumed gas velocities $u_{\rm gas}$ were divided by radial velocities $RV$ as measured from the deepest point of the line profile. The results, summarised in Table\,\ref{t:factorp}, are consistent with the findings from the literature listed above (and more recent results concerning Cepheids by Nardetto et al. \cite{NMMFG07} or Groenewegen \cite{Groen07}). By applying the same method to a hydrostatic MARCS model atmosphere from Aringer (\cite{Aring00}) resembling the star Arcturus (K1.5\,III), we found values reaching up to $p$\,$\approx$\,1.5 (see Sect.\,3.4 of N05).

\begin{table}
\begin{center}
\caption{Conversion factors \textit{p} as derived from the line profiles shown
in Fig.\,\ref{f:dopplershift}. }
\begin{tabular}{ccc}
\hline\hline
$u_{\rm gas}$&$RV$&$p$\\
$[$km\,s$^{-1}$$]$&$[$km\,s$^{-1}$$]$&\\
\hline
5&3.90&1.28\\
10&8.10&1.24\\
20&16.70&1.20\\
30&23.85&1.26\\
40&32.05&1.25\\
\hline
\end{tabular}
\label{t:factorp}
\end{center}
\end{table}  

In some cases, the shapes of observed spectral lines may appear relatively similar to those shown in Fig.\,\ref{f:dopplershift}. One example is given in the right panel of Fig.\,\ref{f:co16mueprofiles-WHyavgl} with the profiles of lower resolution based on model~W. In general, however, the velocity fields in AGB atmospheres are more complex than uniform velocity fields and the resulting line profiles show -- especially for high spectral resolutions -- effects beyond shifts in wavelength. This will be examined below. The estimation of the conversion factor $p$ can also be hindered as illustrated in Sect.\,\ref{s:remarkvels}.

\subsection{Consistent dynamical AGB star spectra}
\label{s:realworld}

\begin{figure}
\resizebox{\hsize}{!}{\includegraphics[clip]{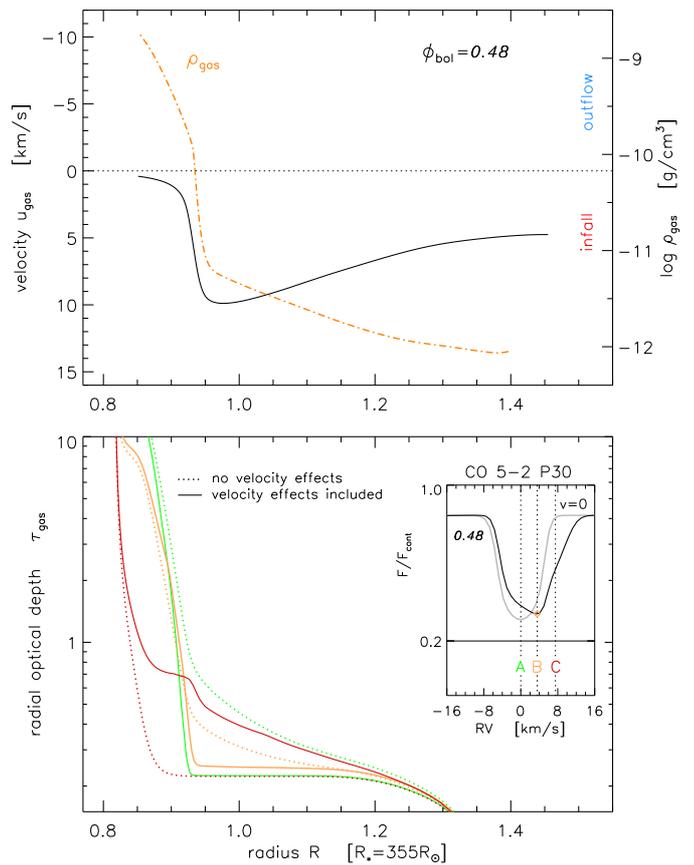}}
\caption{Demonstration of the influence of atmospheric velocities on line
shapes. \textit{Upper panel:} Gas velocities of one selected phase of the
dust-free model~W (black), overplotted is the corresponding density
structure (dash-dotted). \textit{Lower panel:} The insert depicts synthetic CO
$\Delta v$\,=\,3 line profiles  calculated with (black) and without (grey) taking velocities into account for the radiative transfer. The main plot shows the radial optical depth distributions (central beam) for three points A/B/C in the line profile corresponding to different velocity shifts. The correlation is provided by the colour-code.}
\label{f:asymmetry}
\end{figure}

The second example is shown in Fig.\,\ref{f:asymmetry}, where we present a synthetic CO $\Delta v$\,=\,3 line profile based on one selected phase of model~W ($\phi_{\rm bol}$=\textit{0.48}; part of the results in Fig.\,\ref{f:co16mueprofiles-WHyavgl}). The velocity structure of the chosen phase in the upper panel shows infall for all atmospheric points. However, the velocities are quite diverse and not even monotonic, especially in the line forming region of the second overtone CO lines below $\approx$1\,$R_{\star}$ (cf. Sect.\,\ref{s:when-linedoubling}). The smearing of opacity due to varying Doppler-shifts at different depths leads to a clearly asymmetric shape of the absorption line, as shown in the insert of the lower panel. This is in addition illustrated by plots of the radial optical depth at wavelengths corresponding to the line center (A), the line wing (B), and a (pseudo-)continuum point next to the line (C) in the lower panel. The bulk of the opacity in the relevant layers is red-shifted and the deepest point of the spectral line (B) is located at $RV$=\,3.55\,km\,s$^{-1}$ if velocities are taken into account. The layers directly below the emerging shock wave ($<$0.95\,$R_{\star}$) contribute most to the wavelengths marked with (A) and (B). In addition, there is a strongly enhanced absorption at wavelengths corresponding to $RV$$\approx$8\,km\,s$^{-1}$ (C). This is contributed by layers with log\,$\rho_{\rm gas}$\,$<$\,--11.5\,g\,cm$^{-3}$ (cf. Sect.\,3.3.1 in N05) slightly outwards the step in density at $\approx$0.95\,$R_{\star}$ and leads to the destorted line profile.

For the third example concerning the influence of atmospheric velocities on line profiles we refer to Sects.\,\ref{s:tracemove} and \ref{s:when-linedoubling}, where we discuss the effects of a shock wave propagating through the line forming region of CO $\Delta v$\,=\,3 lines. This may not only result in asymmetric line shapes, but also in the phenomenon of line doubling for certain cases (Fig.\,\ref{f:when-doubling}).

For the fourth example we now turn to low-excitation CO $\Delta v$\,=\,2 lines and the example which can be found in Fig.\,3 of Paper\,I. Compared to the second overtone CO lines, which originate in a relatively narrow depth range, these first overtone CO lines are formed throughout the dust-forming region with contributions from many layers with diverse velocities (Paper\,I). Therefore the observed and the synthetic spectra show complex, multi-component line profiles, as discussed in detail in Papers\,I+II. Furthermore, the redistribution of opacity caused by atmospheric velocity fields may also influence the region of line formation of the strongest component. This was exemplified in Sect.\,5.3 of Paper\,I for the results of the synthetic CO 2--0 R19 line profile for one phase of model~S.

\section{Simulating line profile variations} \label{s:simulating}

It turned out in the past that high-resolution (near-IR) spectrospcopy and the study of line profile variations represents the most promising tool to trace the kinematics in AGB atmosphere (e.g. GH04; their Sects.\,4.8.3.1-2). This is exemplified in Fig.\,\ref{f:complexlineformation} with the help of modelling results.

\subsection{Probing different atmospheric depths}\label{s:probing}

\begin{figure}
\resizebox{\hsize}{!}{\includegraphics[clip]{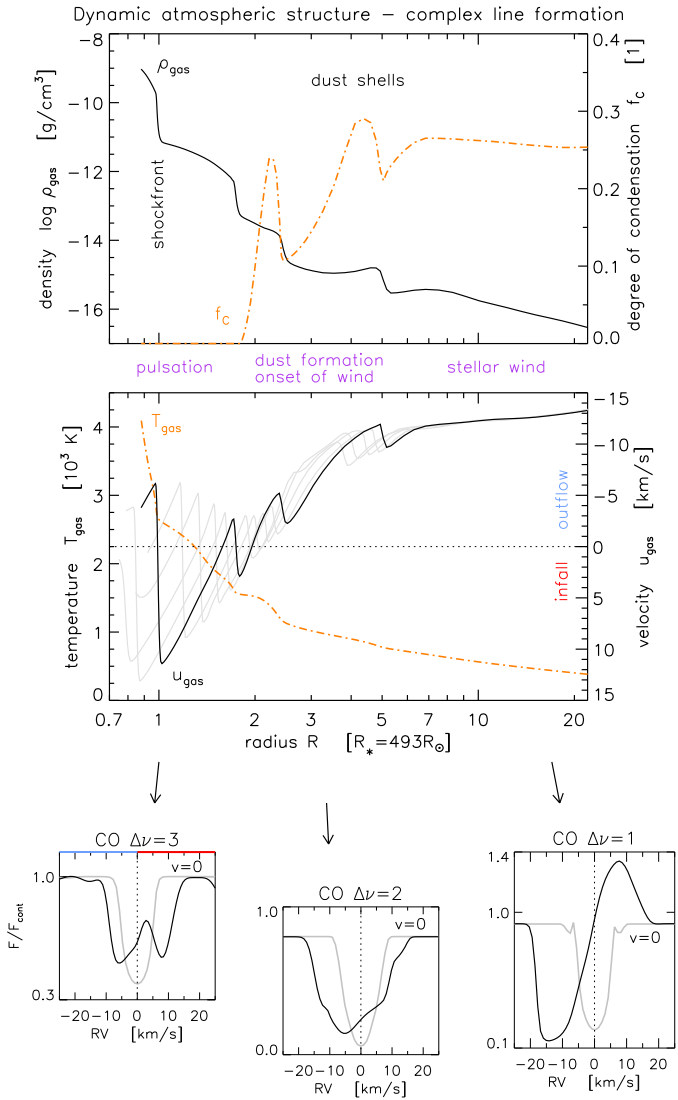}}
\caption{Dynamics within the very extended AGB atmospheres results in complex, time-dependent structures (Sect.\,\ref{s:DMAsused}). This is illustrated in the upper panels, showing the corresponding gas densities, temperatures and velocities as well as the degree of dust condensation for one phase of model~S. Together with the diverse properties of spectral lines (Sect.\,\ref{s:lineformation}) and the significant influence of velocity fields (Sect.\,\ref{s:velocityeffects}), this leads to a complex line formation process. Thus, the synthetic line profiles of selected CO lines of the $\Delta v$\,=\,3\,/\,2\,/\,1 vibration-rotation bands computed for the respective model differ strongly, depending on the atmospheric regions where the lines are formed. This is demonstrated in the lower panels where typical lines (taken from Paper\,I and Fig.\,\ref{f:co16mueprofiles-HHRvgl}) are plotted for the cases that velocity effects are accounted for in the radiative transfer (black) or not (grey). The resulting profiles of different lines can be used to probe gas velocities at separated atmospheric layers (cf. Sect.\,\ref{s:globalvelfield}, Fig.\,9 in Paper\,I).}
\label{f:complexlineformation}
\end{figure}

In their Sect.\,4.8.1, GH04 pointed out that the atmosphere of an evolved AGB star (pulsating and mass-losing) can be subdivided into three distinctive regions: the deep and dust-free photospheric layers dominated by the pulsating stellar interior \mbox{(zone \textit{A}),} the region of dust formation where the stellar wind is triggered (zone \textit{B}), and finally the region of steady outflow at the inner circumstellar envelope (zone \textit{C}). These differ strongly from the dynamic point of view. Our dynamical model atmospheres show a similar behaviour, as can be recognised in Fig.\,\ref{f:complexlineformation} or Fig.\,\ref{f:massenschalenM}. 

Observational spectroscopic studies (e.g. HHR82) demonstrated that individual spectral lines exhibit quite different line forming regions and can be used to gather information (e.g. velocity fields) concerning the above mentioned zones. Molecular lines of CO in the NIR proved to be especially useful in this context as discussed in Sect.\,2.2 of Paper\,I or Sect.\,1.3.2 of N05. A notable amount of high-resolution spectroscopy data of LPVs was collected during the past thirty years, basically by Hinkle, Lebzelter and collaborators. For a detailed review of the results and a comprehensive compilation of references we refer to Sect.\,1.3 of N05. From these investigations, it is known that second overtone ($\Delta v$\,=\,3), first overtone ($\Delta v$\,=\,2), and fundamental ($\Delta v$\,=\,1) vibration-rotation CO lines roughly probe GH04's zones \textit{A}, \textit{B}, and \textit{C}. Hinkle and collaborators derived excitation temperatures for certain types of lines by curve-of-growth analyses. The estimated temperatures in the regions of line formation for CO $\Delta v$\,=\,3\,/\,2\,/\,1 lines amount to approximately 2000--4500\,/\,800--1200\,/\,300\,K.

In Papers\,I+II we studied the line formation including velocity effects for various molecular features (cf. Table\,\ref{t:chosenlines}) on the basis of the same dynamic model atmospheres as used in this work (Sect.\,\ref{s:modelling}). This led to the conclusion that the models from DMA3 are able to reproduce the finding that various spectral features originate at different geometrical depths within the atmosphere. We estimated the temperatures in the line-forming region of CO $\Delta v$\,=\,3\,/\,2\,/\,1 lines to be about 2200--3500\,/\,800--1500\,/\,350--500\,K (Paper\,I), which agrees well with the values derived from observations. Sampling atmospheric layers comparable to the three zones specified by GH04, the synthetic line profiles of the different CO lines show quite different shapes if velocity effects are taken into account in the spectral synthesis. Figure\,\ref{f:complexlineformation} illustrates the strong influence of atmospheric velocities at different depths with characteristic CO line profiles from our model calculations. It also demonstrates how the gas velocities convert into radial velocities, measured by the wavelength shifts of the deepest points of the complex line profiles.

\subsection{Tracing atmospheric movements over time}\label{s:tracemove}

Repeatedly studying one selected line profile in (observed) spectra and deriving RVs allows us to investigate the movement of the atmospheric layers where the respective feature is formed. 

\begin{figure}
\resizebox{\hsize}{!}{\includegraphics[clip]{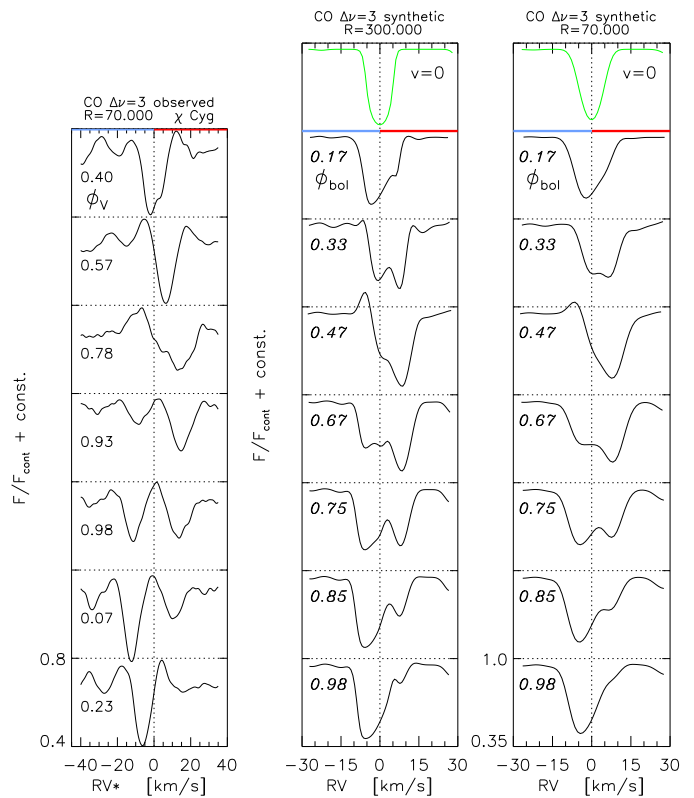}}
\caption{\textit{Left:} Time series of average line profiles of CO $\Delta v$\,=\,3 lines in FTS spectra of $\chi$\,Cyg (cf. Fig.\,1 in HHR82), taken from Lebzelter et~al. (\cite{LebHA01}) who analysed the radial velocities. Observed heliocentric velocities were converted to systemic velocities $RV*$ of the star assuming a \textit{CMRV}$_{\rm heliocent}$ of \mbox{--7.5\,km$\cdot$s$^{-1}$} (HHR82). \textit{Centre and right:} Synthetic second overtone CO line profiles (5--2 P30) seen with spectral resolutions of 300\,000 and 70\,000 for selected phases during one lightcycle, calculated on the basis of model~S.}
\label{f:co16mueprofiles-HHRvgl}
\end{figure}

A prominent example are the often used second overtone vibration-rotation CO lines, which can be found in NIR spectra at $\lambda$\,$\approx$\,1.6\,$\mu$m. As they are well suited to sample the inner photospheric layers where the movements are ruled by the pulsation, CO $\Delta v$\,=\,3 lines show a quite typical behaviour (HHR82) for Mira variables.\footnote{Note that there is a fundamental difference between Miras and SRVs, which is the topic of Sect.\,\ref{s:COdv3SRVs}.} This is exemplified in the left panel of Fig.\,\ref{f:co16mueprofiles-HHRvgl} with a time-series of line profiles (average of 10-20 unblended lines) of the S-type Mira $\chi$\,Cyg for various phases $\phi_{\rm v}$ during the lightcycle. Repeating in the same way every pulsation period, the following characteristic pattern is found.\footnote{The simplified picture of independently evolving components may be misleading to some extent, though, as the spectral features are formed over a whole region in depth and the line profiles (sum of different contributions) can appear more complex (e.g. middle panel of Fig.\,\ref{f:co16mueprofiles-HHRvgl}). Still, it can help to understand the scenario and for the certain phases of line-doubling the velocity jump due to the shock front results in two well-pronounced absorption components.} A blue-shifted component becomes visible before light maximum ($\phi_{\rm v}$$\approx$0.8). With increasing intensity it moves towards the red, crossing $RV$=0 at around $\phi_{\rm v}$$\approx$0.4. Then the component becomes weaker again and disappears red-shifted shortly after the next light maximum ($\phi_{\rm v}$$\approx$0.1). This behaviour leads to the characteristic discontinuous, S-shaped RV curve as shown for example in Fig.\,\ref{f:rv-CO-dv3-allMiras}. The occurrence of line doubling at phases around the visual maximum is usually interpreted by shock waves passing through the line-forming region (Fig.\,1 of Alvarez et al. \cite{AJPGF00}, HHR82, N05). Based on a thorough analysis of these CO $\Delta v$\,=\,3 lines (RVs, excitation temperatures, column densities), HHR82 were able to derive a quite detailed picture of the photospheric structure being affected by the propagating shock wave as well as of the properties of the shock front itself (emergence, progress, velocities, etc.) in their Sect.\,VIa.

In Fig.\,\ref{f:co16mueprofiles-HHRvgl} we compare our modelling results (C-rich model) with observed spectra of the well studied S-type Mira $\chi$\,Cyg, as the spectroscopic monitoring of such CO $\Delta v$\,=\,3 lines for \mbox{C-Miras} is hampered by several problems (severe contamination by mainly CN and C$_2$ in the H-band, availability of targets and instruments for long time series, etc.). Assuming that the principal behaviour should be similar for all spectral types (because of the unique role CO; cf. Sect.\,4.4.5 in GH04), such a comparison seems justifiable. In contrast to the results presented in Paper\,I, we neglected the contribution of other molecular opacity sources (C-bearing molecules, see Paper\,I for details) for the synthetic line profiles shown in Fig.\,\ref{f:co16mueprofiles-HHRvgl} for an easier comparison. As described in detail in Papers\,I+II, we succeeded to reproduce the observed fact that CO $\Delta v$\,=\,3 lines trace the inner pulsating layers of the atmosphere by our modelling approach. Nevertheless, there remained some shortcomings concerning the synthetic line profiles based on model~S. For example, the components of doubled lines are not as clearly separated as in the observations. We also find quite complicated line shapes (at least for the highest spectral resolutions) during phases where observations show a transition of one component from blue- to red-shifts (e.g. phases $\phi_{\rm v}$=0.57 and $\phi_{\rm bol}$=\textit{0.33} in Fig.\,\ref{f:co16mueprofiles-HHRvgl}). This was the motivation for further efforts, the outcome of which will be discussed in Sect.\,\ref{s:realistic}.

\section{Towards realistic models for Miras} \label{s:realistic}

In Papers\,I+II we were able to show that the used dynamic model atmospheres allow us to qualitatively reproduce the characteristic behaviour of spectral features sampling different atmospheric regions by consistent calculations. However, we also pointed out (Sect.\,6 in Paper\,II) the shortcomings of the used model~S and listed aims for subsequent modelling efforts (reasonable velocity amplitudes in the deep photosphere or global velocity variations in quantitative agreement with observed stars). Below, we try to vary the input parameters of the model (Table\,\ref{t:dmaparameters}) to get even more realistic atmospheric structures. The results of this preliminary parameter study is the topic of the next two subsections.

\subsection{The pulsating layers (CO $\Delta$v\,=\,3 and CN lines)}
\label{s:largerampl}

The first and foremost aim for the fine-tuning of model parameters was a dynamic model that exhibits an amplitude of the gas velocities within the pulsating layers that is closer to what is found in observations. 

\begin{figure}
\resizebox{\hsize}{!}{\includegraphics[clip]{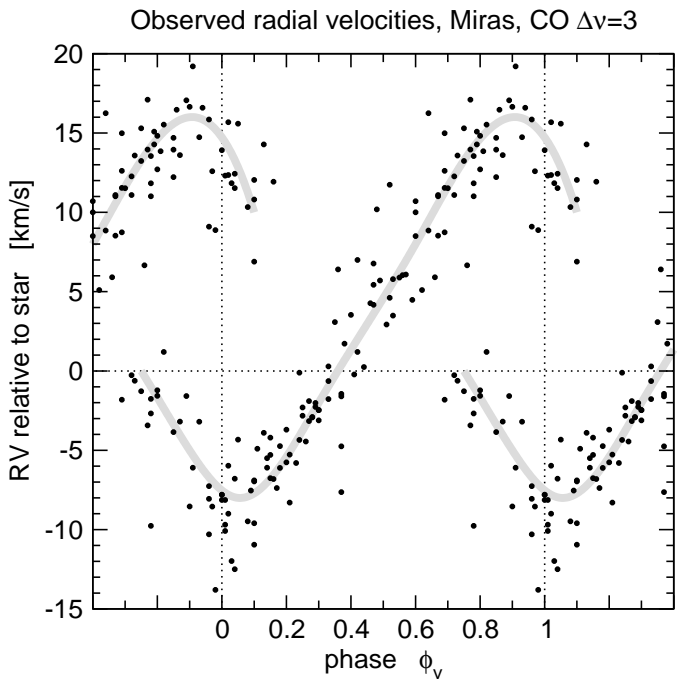}}
\caption{Compilation of RVs derived from second overtone CO lines in FTS
spectra of a large sample of Miras, demonstrating the rather uniform
behaviour. Observed heliocentric RVs are converted to systemic velocities by
the respective \textit{CMRV} of each object. The grey line was fit to guide
the eye. Data adopted from Lebzelter \& Hinkle (\cite{LebzH02a}).}
\label{f:rv-CO-dv3-allMiras}
\end{figure}

Hinkle et al. (\cite{HinSH84}; HSH84) state that though Miras are "clearly individuals" concerning the behaviour of CO $\Delta v$\,=\,3 lines (differences in phases of line-doubling, range and phase-dependency of $T_{\rm exc}$, shapes and scales of column densities, differences in total RV amplitudes), they still share some characteristic features. The most important one is the behaviour of the line profiles as sketched in Sect.\,\ref{s:tracemove} and Fig.\,\ref{f:co16mueprofiles-HHRvgl}, resulting in the common shape of the velocity variations. Lebzelter \& Hinkle (\cite{LebzH02a}) presented a compilation of measured RVs for most of the Miras studied at that time (by Hinkle, Lebzelter and collaborators) in one plot, which has been adopted here in Fig.\,\ref{f:rv-CO-dv3-allMiras}. It appears that Miras have a rather universal RV curve with discontinuities around maximum phases and a part where velocity increases linearly from blue- to red-shifts (negative to positive RVs) through minimum light. The RV amplitude (difference between minimum and maximum values) amounts to \mbox{$\Delta RV$$\approx$20--30\,km$\cdot$s$^{-1}$} for all objects. This uniform picture seems to be valid over a wide range of effective temperatures, periods, (probably) metallicities and for different atmospheric chemistries. The last one was actually found only for \mbox{M-/S-type} stars, as measuring CO $\Delta v$\,=\,3 lines in carbon star spectra proves to be challenging (Barnbaum \& Hinkle \cite{BarnH95}).

\begin{figure}
\resizebox{\hsize}{!}{\includegraphics[clip]{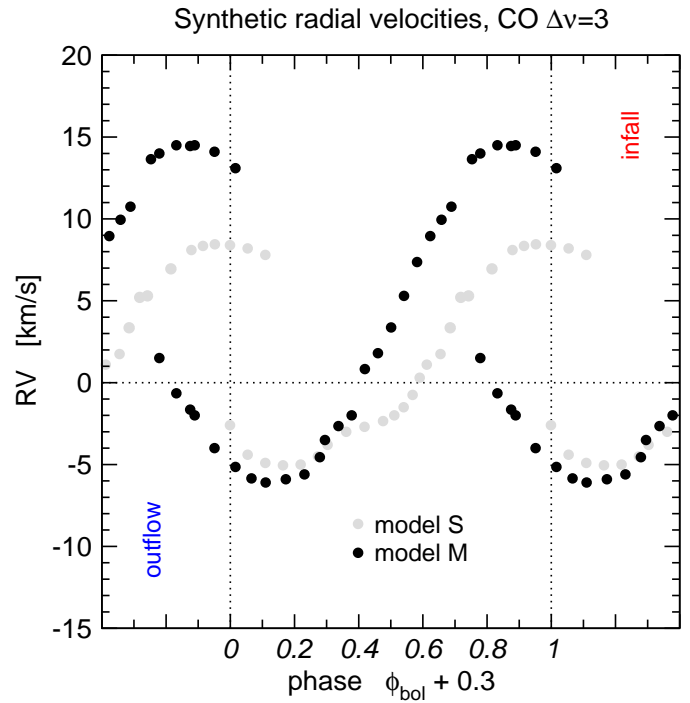}}
\caption{Radial velocities derived from  synthetic line profiles (spectral resolutions of $R$\,=\,70\,000) of second overtone CO lines calculated on the basis of model~S (Fig.\,\ref{f:co16mueprofiles-HHRvgl}, cf. Figs.\,2+9 in Paper\,II) as well as the corresponding ones for the new model~M. An arbitrary shift of 0.3 in phase $\phi_{\rm bol}$ was applied for easier comparison with the observational results in Fig.\,\ref{f:rv-CO-dv3-allMiras}, which are given in $\phi_{\rm v}$. The typical S-shaped, discontinuous RV curve as well as the amplitude $\Delta RV$ found in observations (Fig.\,\ref{f:rv-CO-dv3-allMiras}) can be reproduced with model~M.}
\label{f:rv-CO-dv3-modelsMS}
\end{figure}

As it is a common feature of all Miras, realistic model calculations need to reproduce the discussed RV curve of Fig.\,\ref{f:rv-CO-dv3-allMiras}. Our original model~S was able to fulfill this qualitatively (cf. Fig.\,\ref{f:rv-CO-dv3-modelsMS}). However, the step in gas velocity at the location of the shock was too small. Therefore, the splitting of CO $\Delta v$\,=\,3 lines appeared too weak ($\phi_{\rm bol}$=\textit{0.75} in Fig.\,\ref{f:co16mueprofiles-HHRvgl}) and the derived velocity amplitude was too low ($\Delta RV$$\approx$14\,km$\cdot$s$^{-1}$). Model~M, the result of our small parameter study, was tuned to fit this aspect. Among the several modified parameters the increased piston velocity amplitude $\Delta u_{\rm p}$ is of particular importance. This led to a more realistic variability of the atmospheric structure in the inner parts. The difference between post-shock outflow velocity and pre-shock infall velocity can reach values close to the estimate of $\approx$34\,km$\cdot$s$^{-1}$ given by Scholz \& Wood (\cite{SchoW00}; summarised in Paper\,II), as can be seen in Fig.\,\ref{f:structure} for $\phi_{\rm bol}$=\textit{0.72}. The synthetic CO $\Delta v$\,=\,3 line profiles look similar to those shown in Fig.\,\ref{f:co16mueprofiles-HHRvgl}, the RV curve -- shown in Fig.\,\ref{f:rv-CO-dv3-modelsMS} -- appears much more like the observational results in Fig.\,\ref{f:rv-CO-dv3-allMiras}, though.\footnote{Keeping in mind that the line profiles as well as the RVs may become more complex for the highest resolution of $R$\,=\,300\,000 (see Papers\,I+II), we will use only the RV results of the lower resolution of $R$\,=\,70\,000 to be compatible to observational results.} Model~M is able to reproduce this fundamental characteristic of Miras even quantitatively. Although we needed to apply an arbitrary phase shift of $\Delta\phi$=0.3 to the synthetic RV curve in Fig.\,\ref{f:rv-CO-dv3-modelsMS} to achieve agreement in phase with the observations in Fig.\,\ref{f:rv-CO-dv3-allMiras}, the discontinuous RV curve resembles observations concerning the S-shape, the line doubling interval, the zero-crossing phase, the asymmetry w.r.t. $RV$\,=\,0 (infall velocities larger than outflow velocities as observed for most Miras), and -- mainly -- the velocity amplitude $\Delta RV$ of $\approx$21\,km\,s$^{-1}$.

\begin{figure}
\resizebox{\hsize}{!}{\includegraphics[clip]{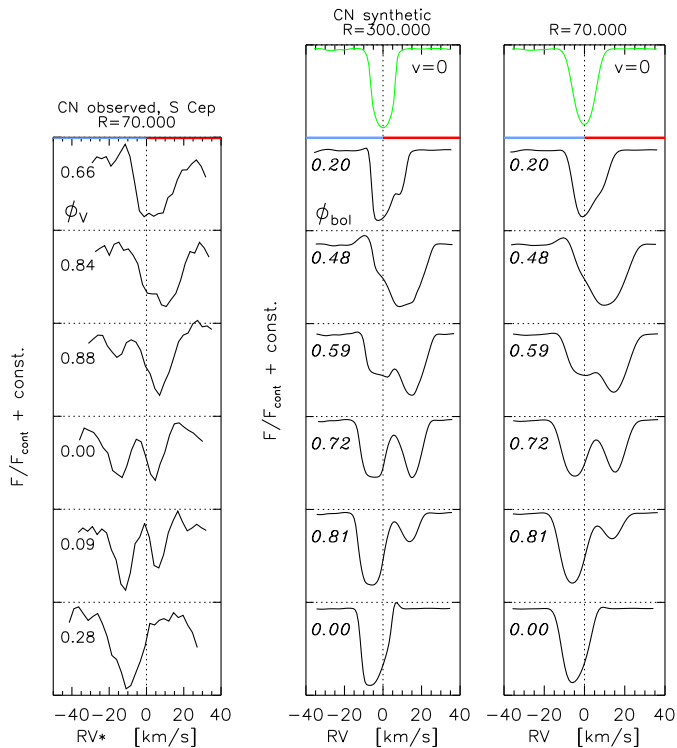}}
\caption{\textit{Left:} CN $\Delta v$\,=\,--2 lines profiles (1--3
P$_2$38.5) in the FTS spectra of S\,Cep observed by HB96 for several phases (cf. Fig\,3 in Paper\,II). Observed heliocentric velocities were converted to systemic velocities $RV*$ of the star assuming a \textit{CMRV}$_{\rm heliocent}$ of --26.7\,km$\cdot$s$^{-1}$ (HB96). \textit{Centre and right:} Sequence of synthetic profiles of similar CN lines (1--3 Q$_2$4.5) for different spectral resolutions of 300\,000 and 70\,000 throughout a lightcycle computed by using model~M.}
\label{f:cn2mueprofiles-modelM}
\end{figure}

Hinkle \& Barnbaum (\cite{HinkB96}; HB96) demonstrated how to circumvent the difficulties in observing CO $\Delta v$\,=\,3 lines in C star spectra by investigating spectral features of CN in the NIR. Also sampling the inner photosphere of a Mira, they show a similar behaviour as second overtone CO lines (Sect.\,\ref{s:tracemove}, Fig.\,\ref{f:co16mueprofiles-HHRvgl}). This can be recognised from the line profiles (left panel of Fig.\,\ref{f:cn2mueprofiles-modelM}), which repeat in the same way every lightcycle and the deriveable RV curves (right panel of Fig.\,\ref{f:rvs-chicyg-scep}). On the basis of the improved model~M, we synthesised CN line profiles following the approach of Paper\,I. The results of the modelling for selected phases are shown in the middle and right panel of Fig.\,\ref{f:cn2mueprofiles-modelM}. It becomes immediately apparent that compared to the profiles calculated with the previous model~S (see Fig.\,5 in Paper\,I), the new line profiles reproduce observations much more realistically. Although substructures may be identified at the highest resolution of $R$=300\,000 (as for model~S), the characteristic pattern of lines sampling the deep pulsating layers is clearly visible, especially when rebinned down to the lower spectral resolution of 70\,000. The transition of one component from blue- to red-shift ($\phi_{\rm bol}$$\approx$\textit{0.9}\,$\leadsto$\,\textit{0.5}) is much smoother. Pronounced line-doubling is found due to the more extreme velocity gradient across the shockfront. The splitting into two well separated components during certain phases can be recognised in Fig.\,\ref{f:cn2mueprofiles-modelM} for $\phi_{\rm bol}$=\textit{0.72} (compare the corresponding velocity structure of the model in Fig.\,\ref{f:structure}). The more realistic atmospheric structure of model~M is also reflected in the RVs derived from the synthetic CN lines. The resulting discontinuous RV curve -- included in Fig.\,\ref{f:rvsmodelM} -- shows the same S-shape\footnote{Interesting is that this is not clearly recognisable in observed CN velocities of S\,Cep as shown in the upper right panel of the same figure. There the RV curve appears more like a straight line. One may suspect that the cross-correlation technique, applied by HB96 to obtain these velocities from the FTS spectra, is not able to resolve the weak components, as they appear shortly before light maximum and disappear shortly after the next maximum.} as the corresponding one from CO $\Delta v$\,=\,3 lines. The small phase shift between these two RV curves suggests that the line-forming region of CN lines lies somewhat further away from the centre of the star and the shock wave passes through at a later phase of the pulsation cycle.\footnote{The mean difference between the RV curves of the two types of lines ($\Delta\phi_{\rm bol}$$\approx$\textit{0.1}) together with an estimate of the propagation velocity of the shock wave ($u_{\rm front}$\,$\approx$\,10.16\,km\,s$^{-1}$; see Sect.\,\ref{s:remarkvels}) would result in a difference $\Delta R$ of approximately 0.15\,$R_\star$\,=\,61.8\,$R_{\odot}$\,=\,0.288\,\textit{AU}.} The velocity amplitude for the synthetic CN lines of $\Delta RV$$\approx$22\,km\,s$^{-1}$ resembles nicely the value of 22.3 found for S\,Cep by HB96 (previously, we found $\approx$13.5\,km\,s$^{-1}$ for model~S). The above mentioned arbitrary phase shift of $\Delta\phi$=0.3, which is needed to align the synthetic RV curve with the observed one in Fig.\,\ref{f:rvs-chicyg-scep}, appears plausible here as well. It is remarkable that the RV curve of S\,Cep is extending to more negative velocities w.r.t. the CMRV, while the modelling results are -- in agreement with observational as well as synthetic CO $\Delta v$\,=\,3 velocities -- shifted to more positive values.

Note that the stronger piston of the improved model~M ($\Delta u_{\rm p}$=6\,km\,s$^{-1}$ instead of 4\,km\,s$^{-1}$ for \mbox{model S;} cf. Table\,\ref{t:dmaparameters}) results only in slightly higher outflow velocities, while the maximum infall velocities are increased considerably, as can be seen in Fig.\,\ref{f:rv-CO-dv3-modelsMS}. The reason for this effect is the larger surface gravity (log\,$g_\star$) of \mbox{model~M} compared to model~S. Thus, the increase in total velocity amplitude $\Delta RV$ is mostly due to the faster infalling material.

\subsection{The global velocity field}\label{s:globalvelfield}

\begin{figure*}
\centering
\includegraphics[width=17cm]{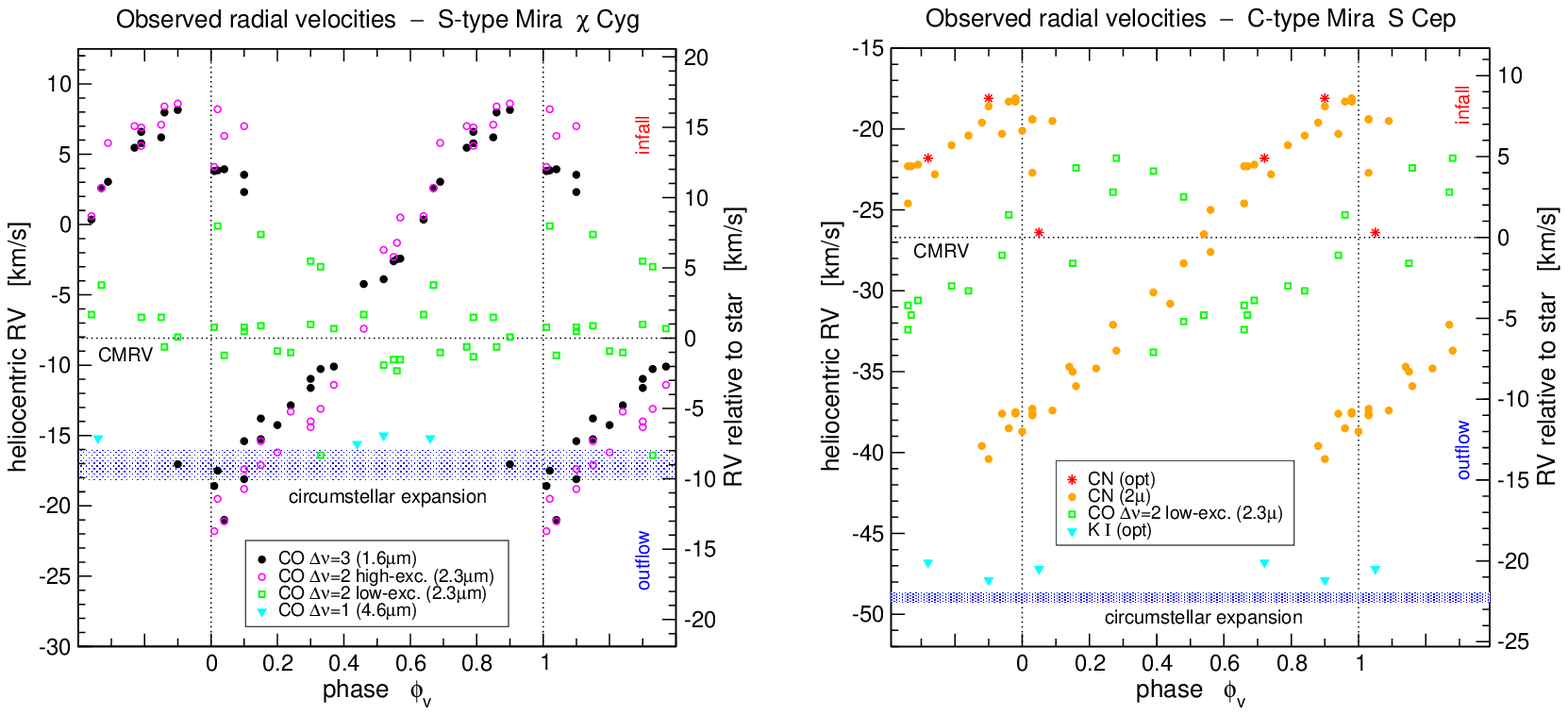}
\caption{Radial velocities derived from (components of) observed or synthetic line profiles providing information on the gas velocities in the respective line-forming region. 
\newline
\textit{Left:} Radial velocities vs. visual phase during a lightcycle ($\phi_{\rm v}$=0 corresponds to phases of maximum light in the visual) for several types of CO lines in observations of $\chi$\,Cyg. The data are adopted from HHR82. Observed heliocentric velocities are converted to systemic velocities of the star assuming a \textit{CMRV}$_{\rm heliocent}$ of --8.08\,km$\cdot$s$^{-1}$, which results from applying the general conversion \textit{RV}$_{\rm heliocent}$=\textit{RV}$_{\rm LSR}$--17.98 (HHR82) to the \textit{CMRV}$_{\rm LSR}$ given by the more recent study of KYL98. Also marked (hatched blue) is the range of expansion velocities of the outflowing circumstellar material as measured by several publications on radio observations (DRM78, KYL98, NKL98, BGP89, LFO93, RSOL06).
\newline
\textit{Right:} Radial velocities vs. visual phase for different lines in observations of S\,Cep. The data are adopted from HB96 and Barnbaum (\cite{Barnb92b}). Observed heliocentric velocities are converted to systemic velocities of the star assuming a \textit{CMRV}$_{\rm heliocent}$ of --26.65\,km$\cdot$s$^{-1}$, which results from applying the general conversion \textit{RV}$_{\rm heliocent}$=\textit{RV}$_{\rm LSR}$--11.65 (HB96) to the \textit{CMRV}$_{\rm LSR}$ given by BGP89. Again, the range of expansion velocities of the outflowing circumstellar material as measured by several publications on radio observations (B92a, NKL98, BGP89, LFO93, OLN98, OEG93, RSOL06) is marked (hatched blue).}
\label{f:rvs-chicyg-scep}
\end{figure*}

Simultaneous spectroscopic monitoring of various spectral features originating in different layers and for several instances of time (e.g. during a pulsation period) enables us to trace the evolution of the global velocity field throughout the outer layers of an AGB star and thereby the mass loss process.

Extensive time series IR spectrocopy, which is needed to characterise the overall atmospheric dynamics, is available only for a few stars (limited availability of high-resolution IR spectrographs, demanding observations over years due to the long periods). Investigated with sufficient coverage in phase and spectral range (i.e. line types) were mainly the O-rich Mira R\,Leo, the S-type Mira $\chi$\,Cyg, and the C-rich Mira S\,Cep (N05). HHR82 presented a fundamental study on $\chi$\,Cyg. Their time series of NIR spectra and RVs derived from different spectral features revealed remarkable details about the atmospheric motions of this star.\footnote{$\chi$\,Cyg appears to be the most extensively studied LPV from the kinematics point of view, at all, as outlined in Sect.\,1.3.4 of N05.} Although dealing with an S-type star, the results can be considered to be representative for Miras in general due to the singular role of the CO molecule (e.g. Sect.\,4.4.5 in GH04) and the resulting presence in atmospheres of objects of all spectral types. Obtained radial velocities from HHR82 are compiled and plotted in the left panel of Fig.\,\ref{f:rvs-chicyg-scep}, supplemented with measurements of the expansion velocity of the circumstellar envelope (CSE) for comparison. There is only one C-type Mira, namely S\,Cep, for which a similar plot can be produced. The right panel of Fig.\,\ref{f:rvs-chicyg-scep} shows the corresponding RVs compiled from HB96 and Barnbaum (\cite{Barnb92b}), as well as again the range of CSE outflow velocities coming from radio observations of molecular emission lines.\footnote{A similar compilation of RV data for the M-type Mira R\,Leo (collected by K. Hinkle and collaborators) can be found in Fig.\,1.19 of N05.} While CO $\Delta v$\,=\,3, CN and CO $\Delta v$\,=\,2 high-excitation lines sample the regularly pulsating layers of the deep photosphere, CO $\Delta v$\,=\,2 low-excitation lines are formed in the dust-forming layers. CO $\Delta v$\,=\,1 and selected atomic lines (e.g. K\,I at 7698.96\,\AA{}) probe the outermost wind region.

\begin{figure}
\resizebox{\hsize}{!}{\includegraphics[clip]{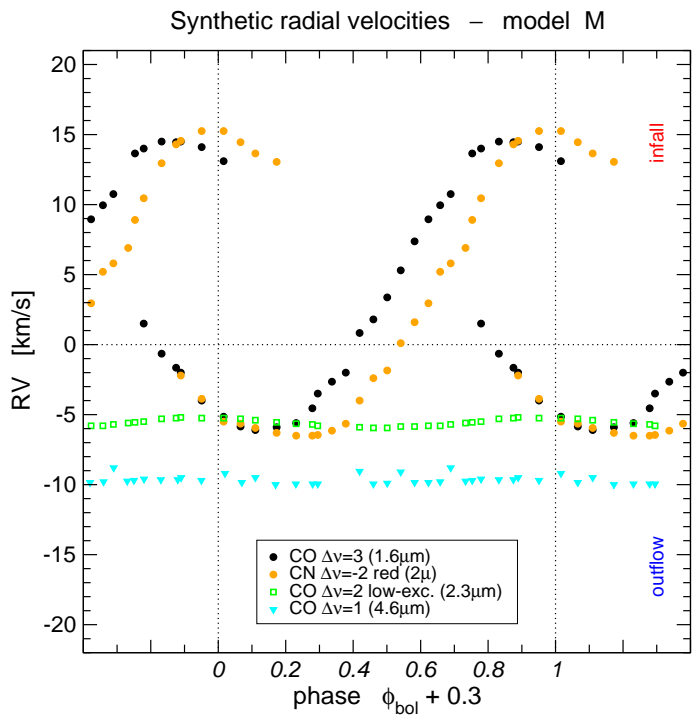}}
\caption{Radial velocities derived from various synthetic line profiles (spectral resolutions of $R$=70\,000) computed on the basis of model~M. An arbitrary shift of 0.3 in phase was applied for easier comparison with the observational results in Fig.\,\ref{f:rvs-chicyg-scep}.}
\label{f:rvsmodelM}
\end{figure}

In Fig.\,\ref{f:rvsmodelM} the RV results of our line profile modelling based on model~M are shown, aiming also for a global mapping of the dynamics throughout the atmosphere. As already described in detail in the previous section, the new model~M proves to be rather realistic concerning the velocity variations in the deep atmospheric layers (governed by the periodic pulsations) and the resulting line profile variations of molecular features originating there (i.e. CO $\Delta v$\,=\,3, CN).

Observed low-excitation CO $\Delta v$\,=\,2 lines were found to probe the dust-forming region (Paper\,I, Sect.\,\ref{s:tracemove}) and show broadened, asymmetric line shapes (e.g. Fig.\,3 in HB96). The multi-component profiles appear to be blends of a few contributions, which are at the highest resolutions currently available not separable individually though (no pronounced line splitting), and the temporal variations are not related to the lightcycle (cf. Papers\,I+II). Although all investigated Miras have in common that the RV (measured from the deepest point of the main component) variations of these lines are not at all periodic (at least on the time scales of the observations), the behaviour can be somewhat different depending on the object considered. For $\chi$\,Cyg the RVs of CO $\Delta v$\,=\,2 low-excitation lines always stay close to the CMRV (Fig.\,\ref{f:rvs-chicyg-scep}). For other stars, as for example R\,Vir (Lebzelter et al. \cite{LebHH99}) or S\,Cep (Fig.\,\ref{f:rvs-chicyg-scep}), the velocities show moderate variations around the CMRV  with amplitudes of up to $\approx$15\,km$\cdot$s$^{-1}$ (in any case lower than the amplitudes of CO $\Delta v$\,=\,3 lines though). Line profiles of the rather special star IRC+10216 (dense and optically thick dust shell resulting from pronounced mass loss) are clearly blue-shifted indicating only outflowing material (Fig.\,9 of Winters et al. \cite{WiKGS00}). These varying behaviours may result from the fact that velocities in the line-forming layers are rather sensitive to the respective stellar parameters and the wind-acceleration process. In addition, optical depth effects will have some influence. Synthetic CO $\Delta v$\,=\,2 line profiles based on \mbox{Model~M} show -- apart from some temporally varying asymmetries, which can be interpreted as minor photospheric contributions -- a strong main component, which appears blue-shifted at all phases. The slow but quite steady outflow velocity of $\approx$--5\,km$\cdot$s$^{-1}$ (Fig.\,\ref{f:rvsmodelM})  points towards a line formation in layers where the atmospheric material is already accelerated and variations due to shock waves are low (Fig.\,\ref{f:structure}).\footnote{A re-calculation of model~M with an improved version of our RHD code -- increased number of wavelength points for the RT, updated sticking coefficients, grain-size dependent treatment of dust instead of small particle approximation, etc. -- led to a faster (u\,$\approx$\,10--15\,km\,s$^{-1}$) but less dense ($\dot M$\,$\approx$\,10$^{-6}$) wind with a decreased dust content ($f_c$\,$\approx$\,0.1). One may suspect that the decreased gas/dust absorption will lead to CO $\Delta v$\,=\,2 lines originating closer to the centre of the star, where the velocity variations are more pronounced. With the inner regions receiving the same mechanical input and a different wind structure at the same time, the line profiles based on this new model should therefore look more realistic.}
In principle it could also be that the onset of the stellar wind in model~M occurs in a too smooth way and the velocity variations in the dust-forming region are too small compared to a Mira like S\,Cep, which seems unlikely for model~M, though (cf. Fig.\,\ref{f:structure}).

However, general statements on the dynamics in the dust-forming regions of Miras based on high-resolution spectroscopy appear to be not within reach at the moment due to limitations on the observational as well as on the modelling side. For the former we are confronted with a very small number of observed targets (cf. Sect.\,1.3.2.2 in N05), sometimes sparsely sampled in pulsation phase, showing a variety of behaviours (e.g. Lebzelter et al. \cite{LebHH99}). As such low-excitation CO $\Delta v$\,=\,2 lines are formed over a wide radial range within the stellar atmosphere, we face an intricate line formation process (influence of velocity fields, different origins) and, thus, complex line profiles. These are very sensitive to the atmospheric structure ($\rho$-$T$-$u$) and the occuring amount of dust, which are both strongly depending on the properties of the star (stellar parameters, pulsation characteristics). In addition, there may be (phase-dependent) deviations from spherical symmetry (e.g. Woitke \cite{Woitk06a}, Freytag \& H\"ofner \cite{FreyH08}) and various layers of different (projected) velocities contributing to the emerging spectrum. In contrast to the general behaviour of CO $\Delta$v\,=\,3 lines in Mira spectra (Sect.\,\ref{s:largerampl}), a cycle-to-cycle or object-to-object comparison of first overtone lines
does not deliver such a uniform pattern. The individuality of different AGB stars is even more prononunced in the remarkable velocity variations derived from spectral features in the 4\,$\mu$m range (Lebzelter et al. \cite{LebHA01}).
We found similar effects by our modelling, although we only used three models to compute synthetic first overtone CO lines. While models~S and M exhibit a rather steady dust formation and CO $\Delta$v\,=\,2 velocities with almost no temporal variations, things become quite different for the model presented in Sect.\,6.1 of Paper\,II (designated model~S1 in N05). This model shows pronounced cycle-to-cycle variations of dust formation resulting in outwards propagating dust shells (cf. Fig.\,10 in Paper\,II or Fig.\,3.17 in N05), and thus remarkably changing velocity fields in the region of line formation for this type of CO lines. It will need further systematic studies -- covering a large number of observational targets as well as models over a reasonable parameter range -- to end up with a more general picture of the behaviour of CO $\Delta$v\,=\,2 low-excitation lines and its relation to the corresponding parameters of the stars or models.\footnote{Such an investigation aiming at reasonably sampled RV diagrams as in Fig.\,\ref{f:rvs-chicyg-scep} or Fig.\,\ref{f:rvsmodelM} appears rather challenging, though, because of the limited access to suitable spectrographs end the efforts needed on the modelling side.}

Although observational studies of CO $\Delta v$\,=\,1 lines are difficult due to severe telluric absorption, these lines still proved to be useful, as they originate in the cool wind region of evolved red giants. Only very few spectroscopic studies (e.g. Bernat \cite{Berna81}) dealt with these fundamental mode lines, all of them finding clearly blue-shifted spectral lines (also with multiple absorption components or P\,Cygni-type profiles). The results of the small time series of spectra for $\chi$\,Cyg obtained by HHR82 are included in Fig.\,\ref{f:rvs-chicyg-scep}, showing similar blue-shifts for every instance of time. We find the same behaviour for the synthetic line profiles based on the improved model~M. The line shapes look very similar to those based on model~S (Fig.\,8 in Paper\,I), although with some asymmetric distortion resulting in RVs (derived from the deepest point of the absorption), which are higher than the actual outflow velocities of the model. The derived RVs of \mbox{$\approx$--10\,km$\cdot$s$^{-1}$} (Fig.\,\ref{f:rvsmodelM}) compare rather well to the values of $\chi$\,Cyg and other Miras, while the C-type Mira S\,Cep appears to have an outflow velocity higher by a factor of $\approx$2, placing this star on the upper end of the velocity distribution for optical C stars (e.g. Olofsson et al. \cite{OlEGC93}+\cite{Olofs04}, Ramstedt et al. \cite{RaSOL06}). 

Our line profile modelling has now reached a state where we can conclude that the dynamic model atmospheres show global velocity structures that are in general quite close to real pulsating and mass-losing AGB stars. In particular, the characteristic and univsersal behaviour of lines originating in the pulsating inner photosphere (CO $\Delta v$\,=\,3, CN) can now realistically be reproduced by our models. The next step would be to fit in addition the velocities in regions where other types of lines originate (simultaneously with one model for a given object). However, such a tuning of the models to achieve a specific fit for a certain star appears to be difficult. Apart from the problems on the observational side (insufficient determination of stellar parameters like luminosity, mass or C/O; sparsely available RV data from high-resolution spectroscopy) and the questionable relation between dynamic models and observed targets (cf. Sect.\,3 in Paper\,II), the modelling itself sets limits to the tuning efforts. The quite time-consuming process from a given set of model parameters (Table\,\ref{t:dmaparameters}) to an RV diagram (Fig.\,\ref{f:rvsmodelM}) will impede the fitting procedure. Still, the outcomes of line profile studies can provide important constraints (in addition to photometry, low-resolution spectra, interferometry, etc.) for a detailed study of selected AGB stars.

\section{CO $\Delta$v\,=\,3 line profiles of SRVs and Miras} 
\label{s:COdv3SRVs}

\subsection{Observed line profile variations of SRVs}
\label{s:COdv3SRVsobs}

The behaviour of second overtone CO lines in spectra of SRVs is somewhat different from those of Mira variables (Sect.\,\ref{s:tracemove}). Time-series spectroscopy of a few selected objects and results on radial velocities were published by Hinkle et al. (\cite{HinLS97}), Lebzelter (\cite{Lebze99a}) and others (cf. Tables\,1.4 and 1.5 in Nowotny \cite{Nowot05}). It appears that there is no RV curve uniformly shaped for all SRVs as it was found for Miras. The velocity variations with time can be smooth, continuous and regular to different degrees. In a simple approach one would assume that irregular RV variations reflect irregularities in the pulsation of the interior (which are also supposed to be the reason for the irregular elements in the light curves of SRVs) in combination with optical depth effects. The RVs are not strictly periodic, velocity values do not repeat from one period to the next. That only Doppler-shifted profiles but no line doubling can be found at any phase represents the clearest distinction from Miras. Semiregular variables have significantly smaller RV amplitudes (difference between maximum and minimum RV) than Miras, most objects studied show only variations in the range of a few km$\cdot$s$^{-1}$. Only very few exceptions show amplitudes of 10--15\,km$\cdot$s$^{-1}$ (among them W\,Hya, cf. Sect.\,\ref{s:whya}). Hinkle et al. (\cite{HinLS97}) state that all LPVs have random velocity variations of the order of a few km$\cdot$s$^{-1}$ (as possible explanations they list large-scale convective phenomena, atmospheric turbulence, non-radial pulsation, random changes in the periods of the stars, etc.), "masking the intrinsic RV variations" due to pulsation in which we are interested here. As they have the same order of magnitude as the low amplitudes of SRVs caused by pulsation, these random RV variations are more relevant for such stars compared to Miras. From curve of growth analyses Hinkle et al. (\cite{HinLS97}) deduced that CO $\Delta v$=3 lines exhibit roughly constant excitation temperatures over the lightcycle (typically 3300--3400\,K) with variations of less than 200\,K. Conspicuous is -- and this is another important difference to Miras -- that the velocity distributions of the SRVs studied by Hinkle et al. (\cite{HinLS97}) and Lebzelter (\cite{Lebze99a}) are clearly asymmetric w.r.t. the systemic velocity $RV$\,=\,0, but in the opposite way. Negative offsets of several km$\cdot$s$^{-1}$ from the CMRV were found, the second-overtone lines appear blue-shifted for most or even all of the lightcycle. This means that only outflow is observed, while no infalling material is seen in the spectra. The authors discuss possible explanations for this puzzling result. Among several reasons (long secondary periods, etc.), effects of convection were proposed as a solution. From observations of our Sun we know that the interplay between the intensity of granules and velocity fields can result in an overall blue-shift. On the surface of late-type giants large convective cells are expected to occur (e.g. Freytag \& H\"ofner \cite{FreyH08}). This could provide an explanation for velocity asymmetries (convective blue-shift). The question was later resumed by Lebzelter \& Hinkle (\cite{LebzH02b}), who did not observe time series of individual stars, but rather chose a statistical approach by obtaining data at two epochs for a quite substantial sample of SRVs. The resulting velocities, together with some older measurements from time series of individual stars can be found in their Fig.\,2. Again, the majority of the measurements are shifted w.r.t. the CMRV. Nevertheless, Lebzelter \& Hinkle rule out that variable convective cells on the surface are the sole reason for variability in these stars (some measurements show red-shifted lines with $RV$\,$>$\,$CMRV$), indicating that we observe a mixture of pulsation, convection and long period variations still to be clarified. Also temperature effects may play a role (only layers with certain temperatures lead to spectral lines, but may be observable only at certain phases).

Although the asymmetric velocity distribution is not yet fully understood and no typically shaped RV curve for all SRVs was found, Fig.\,2 of Lebzelter \& Hinkle (\cite{LebzH02b}) suggests the following general behaviour (as pointed out by the authors). RV variations have on average amplitudes of 3--4\,km$\cdot$s$^{-1}$ with most positive velocities around phases of light minimum ($\phi_{\rm v}$$\approx$0.5) and most negative values around light maximum ($\phi_{\rm v}$$\approx$0.0). The whole distribution appears blue-shifted by $\approx$1\,km$\cdot$s$^{-1}$. The velocity variations may be rather continuous, and line doubling does not occur.\footnote{although there are few exceptions for both, which reflects the diversity of the objects}

\begin{figure}
\resizebox{\hsize}{!}{\includegraphics[clip]{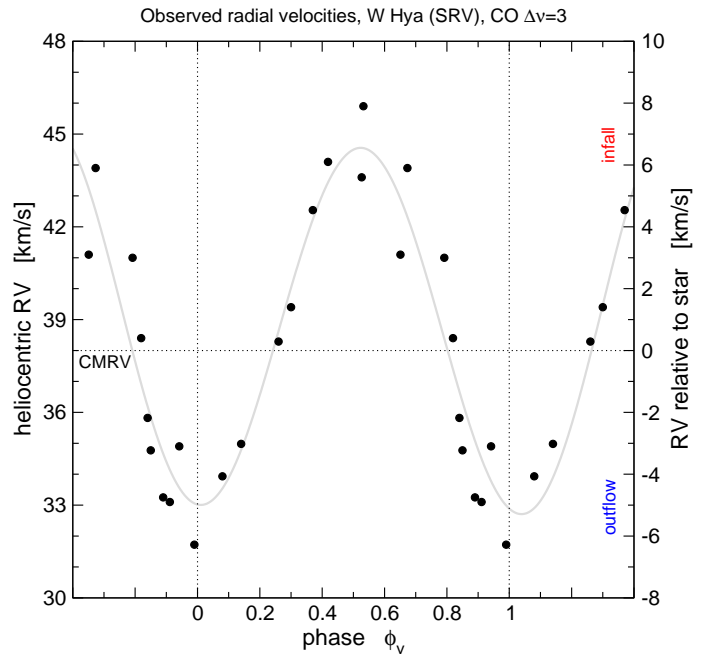}}
\caption{Radial velocities derived from Doppler-shifted CO $\Delta v$\,=\,3 lines in FTS spectra of W\,Hya (left panel of Fig.\,\ref{f:co16mueprofiles-WHyavgl}). Observed heliocentric velocities are converted to systemic velocities of the star assuming a \textit{CMRV}$_{\rm heliocent}$ of 38\,km$\cdot$s$^{-1}$ (Hinkle et al. \cite{HinLS97}). Data adopted from Lebzelter et al. (\cite{LHWJF05a}). The grey line was fit to guide the eye.}
\label{f:RVcurveWHya}
\end{figure}

It is interesting to note that the latter does not necessarily mean that there are no pulsation-induced shock waves. For some SRVs, as e.g. W\,Cyg in Hinkle et al. (\cite{HinLS97}), the hydrogen lines of the Balmer series show up in emission during some light cycles (which is a clear indication for shocks; e.g. Richter et al. \cite{RichW01}+\cite{RWWBS03} and the summarised references therein), even though no doubled lines can be found. We will show by means of model calculations that this behaviour can be reproduced by one of our dynamic model atmospheres.

\subsection{The semiregular variable W\,Hya}
\label{s:whya}

One star, namely the M-type SRV W\,Hydrae, is singled out here, as it will play a role for the following comparison with model~W. Radial velocity variations derived from CO $\Delta$v\,=\,3 lines were analysed by Hinkle et al. (\cite{HinLS97}) and Lebzelter et al. (\cite{LHWJF05a}). This object appears somewhat strange concerning its attributes. Although some irregularities in the shape of the lightcurve and the visual amplitude were found, it appears to be more like a Mira variable from its light variation ($P$\,=\,361$^{\rm d}$, $\Delta V$\,=\,3.9$^{\rm m}$). This is also supported by its location on sequence C (fundamental mode pulsators) in P-L-diagrams (Lebzelter et al. \cite{LHWJF05a}). Based on the limited sample of measurements, Hinkle et~al. speculated that also the velocity variations could be interpreted by a Mira-like behaviour. This was ruled out later by Lebzelter et al. (\cite{LHWJF05a}) using their extended series of spectroscopic observations. From the resulting plot of RVs versus phase, shown in Fig.\,\ref{f:RVcurveWHya}, W\,Hya has clearly to be assigned to the group of (large amplitude) SRVs. It shows a rather periodic, continuous and almost sinusoidal RV curve. No line doubling could be detected. Still, the velocity amplitude of $\approx$15\,km$\cdot$s$^{-1}$ is clearly larger than the average for SRVs. W\,Hya exhibits a well pronounced intrinsic RV variation, perceivable even on top of the (above discussed) random variations. It was suspected by Hinkle et al. (\cite{HinLS97}) that this star could be in a transition from the SRV stage to the Mira stage, an evolutionary scenario that was introduced with the detailed studies of pulsation of AGB stars (cf. Sect.\,\ref{s:intro}). Later, Lebzelter et al. (\cite{LHWJF05a}) concluded that a higher mass might be the reason for the remarkable velocity behaviour (no line doubling and small $\Delta RV$ compared to typical Miras also pulsating in the fundamental mode).

\subsection{Synthetic line profiles and radial velocities}
\label{s:COdv3modelW}

\begin{figure}
\resizebox{\hsize}{!}{\includegraphics[clip]{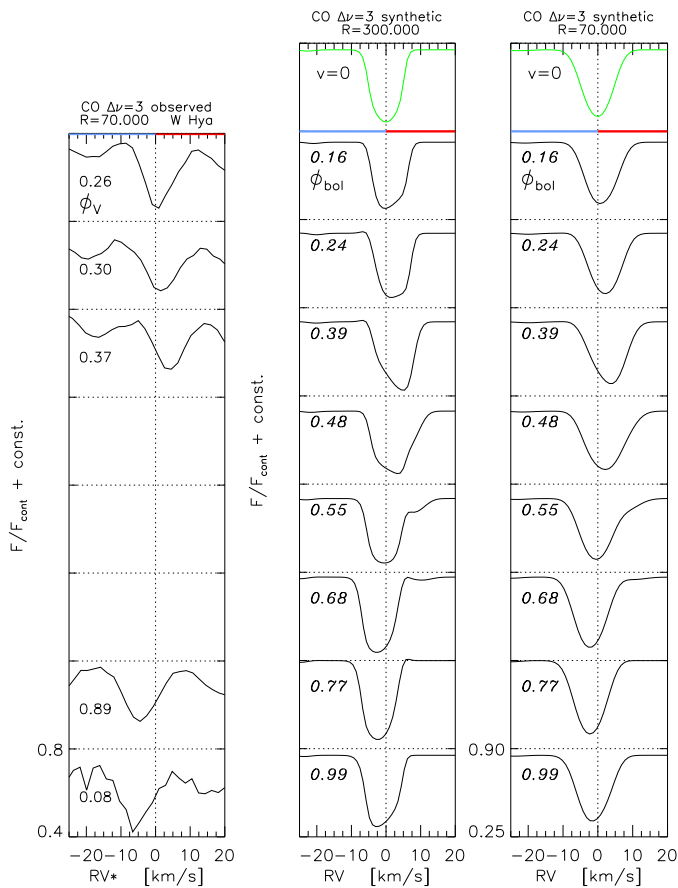}}
\caption{\textit{Left:} Time series of average line profiles of CO $\Delta v$\,=\,3 lines in FTS spectra of the large-amplitude semiregular variable W\,Hya, taken from Lebzelter et al. (\cite{LHWJF05a}) who analysed the radial velocities (cf. Fig.\,\ref{f:RVcurveWHya}). Observed heliocentric velocities were converted to systemic velocities $RV*$ of the star assuming a \textit{CMRV}$_{\rm heliocent}$ of +38\,km$\cdot$s$^{-1}$ (Hinkle et al. \cite{HinLS97}). \textit{Centre and right:} Synthetic second overtone CO line profiles (5--2 P30) at the two different spectral resolutions indicated, calculated on the basis of model~W for selected phases $\phi_{\rm bol}$.}
\label{f:co16mueprofiles-WHyavgl}
\end{figure}

\begin{figure}
\resizebox{\hsize}{!}{\includegraphics[clip]{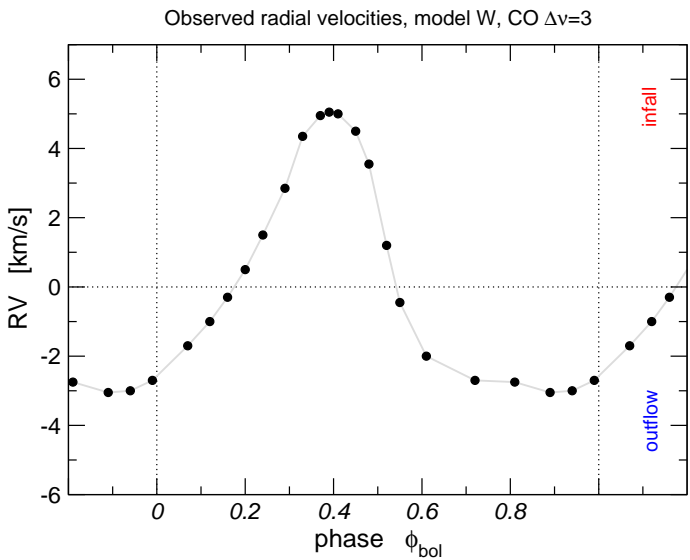}}
\caption{Radial velocities throughout the lightcycle as derived from synthetic CO $\Delta v$\,=\,3 profiles based on model~W (right panel of Fig.\,\ref{f:co16mueprofiles-WHyavgl}).}
\label{f:rv-CO-dv3-modelW}
\end{figure}

While studying synthetic high-resolution spectra on the basis of different dynamic model atmospheres, we found that model~W (described in detail in Sect.\,\ref{s:DMAsused}) is able to reproduce the behaviour of CO $\Delta v$\,=\,3 lines in spectra of SRVs -- particularly of W\,Hya. This model of a pulsating atmosphere was computed with a rather low piston amplitude of $\Delta u_{\rm p}$\,=\,2\,km$\cdot$s$^{-1}$. No dust-formation takes place in the outer layers, and thus no stellar wind can develop. Also, W\,Hya shows only a very low mass loss rate of $\dot M$\,$\approx$\,2$\cdot$10$^{-8}$\,$M_{\odot}$\,yr$^{-1}$ (Hinkle et al. \cite{HinLS97}). It shall however explicitly be noted that the stellar parameters of (the C-rich) model~W were not chosen to resemble (the O-rich star) W\,Hya. Still, a comparison of this one certain aspect -- the RV variations of CO lines -- will be made, as the model reproduces the behaviour characteristic also for all other SRVs observed so far (Sect.\,\ref{s:COdv3SRVsobs}).

The resulting synthetic line profiles for a representative number of phases during the lightcycle are plotted in the two right panels of Fig.\,\ref{f:co16mueprofiles-WHyavgl}. Whenever the corresponding profiles of the observed FTS spectra were available, they are also plotted for comparison in the left panel. The figure demonstrates that the CO $\Delta v$\,=\,3 lines only show a single blue- or red-shifted component as well as rather asymmetric line profiles at highest resolution ($R$\,=\,300\,000) due to the non-monotonic velocity field of the atmosphere (the extreme case of phase $\phi_{\rm bol}$=\textit{0.48} was investigated in detail in Sect.\,\ref{s:realworld} and Fig.\,\ref{f:asymmetry}). RVs derived from the deepest point of the line profiles are drawn in Fig.\,\ref{f:rv-CO-dv3-modelW}. The plot shows that our model~W is able to reproduce the principal behaviour of semiregular variables as sketched in Sect.\,\ref{s:COdv3SRVsobs} (maximum RVs around light minimum, no line doubling). Concerning the asymmetry w.r.t. $RV$\,=\,0 as well as the velocity amplitude ($\Delta RV$\,$\approx$\,8\,km$\cdot$s$^{-1}$ for model~W), the synthetic RV curve is more similar to the observed data of W\,Hya in Fig.\,\ref{f:RVcurveWHya} than to the general behaviour of SRVs shown in Fig.\,2 of Lebzelter \& Hinkle (\cite{LebzH02b}), though. Only a small difference in phase may be suspected, so that the bolometric phases of the model lag behind the visual phases of the observations by 0.1--0.2.

\subsection{Shock waves and the occurrence of line doubling}
\label{s:when-linedoubling}

\begin{figure*}
\centering
\includegraphics[width=17cm,clip]{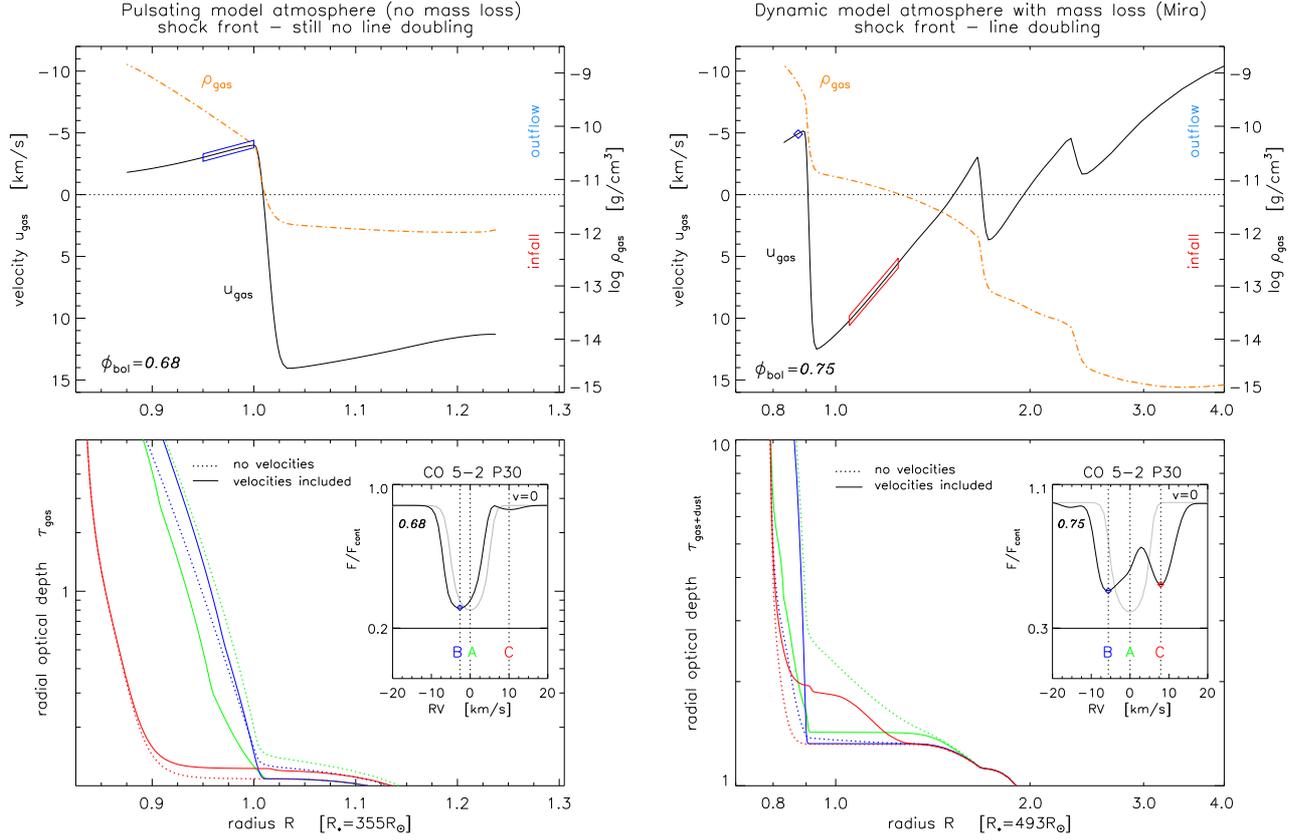}
\caption{Illustration of the fact that the occurrence of a shock wave within the atmosphere does not necessarily lead to line doubling in spectra containing CO $\Delta v$\,=\,3 lines. Demonstrated by representative phases of the dust-free model~W (left) and the wind model~S (right). In the upper panels the atmospheric structures are plotted. The lower panels show the radial optical depth distributions at certain wavelength / RV points as marked with A/B/C in the insets containing the synthetic line profiles (with and without taking velocity effects into account for the radiative transfer). The correlation is provided by the colour-code. See text in Sect.\,\ref{s:when-linedoubling} for details.}
\label{f:when-doubling}
\end{figure*}

As discussed in Sect.\,\ref{s:COdv3SRVsobs}, observational studies of semiregular variables -- in contrast to Miras -- did not find doubled CO $\Delta v$\,=\,3 lines for any phase during the light cycle. This behaviour is somewhat astonishing as one would generally expect shock waves in pulsating atmospheres, and also Balmer emission lines were observed in spectra of selected SRVs without CO line doubling. However, we found the same result for line profiles based on model~W as discussed in the previous section. In Fig.\,\ref{f:when-doubling} the puzzling fact shall be explored with the help of our modelling tools. For selected phases of model~W (left panels) and model~S (right panels), we present in this figure the atmospheric structure (gas velocities and the corresponding densities) together with the resulting CO line profiles and the radial optical depth at significant wavelength points within the synthetic spectra (A/B/C).

For the model without mass loss in the left panels, a shock front can be seen in the velocity structure around 1\,$R_\star$. Nevertheless, only a blue-shifted and slightly asymmetric component with $RV$=\,--2.65\,km$\cdot$s$^{-1}$ (B) from behind the front shows up in the line profile. Practically no red-shifted component is formed. The very weak one, which can be vaguely discerned in the synthetic spectrum at $RV$$\approx$\,10\,km$\cdot$s$^{-1}$ (C), would certainly not be recognisable in observed spectra. The reason for this is the sharp drop in density in front of the shock wave where the necessary gas velocities would be available. The optical depths at wavelength points redder than the rest wavelength of the respective line are not large enough so that a component would become visible in the spectrum.\footnote{Note: the optical depths at every wavelength\,/\,RV point of the spectrum increases dramatically at the inner edge of the dynamical model and further on within the hydrostatic extension (see Sect.\,\ref{s:lineformation}) because of the increasing continuum opacities, though.} Although model~W shows infall velocities of up to $\approx$\,14\,km$\cdot$s$^{-1}$, these are not translated into the spectrum as the related outer layers do not contribute to the formation of the spectral line. As discussed in Sect.\,3.3.1 of N05, the gas densities of the model needed to form an individual component of the used CO $\Delta v$\,=\,3 line have to be higher than \mbox{log\,$\rho_{\rm gas}$\,$\approx$\,--11.5\,g$\cdot$cm$^{-3}$,} which is not met for the phase of model~W displayed in Fig.\,\ref{f:when-doubling}.

However, this requirement is fulfilled for the phase of model~S shown in the right panels of the figure. Due to the formation of dust and the resulting wind, the atmospheric structure differs somewhat in the region around 1\,$R_\star$, which is relevant for the line formation of CO $\Delta v$\,=\,3 lines. The layers in front of the innermost shock front show smaller velocities compared to model~W, but the densities are enhanced by roughly one order of magnitude. Therefore also the infalling material contributes to the spectrum formation. The non-zero velocities of the model result in a redistribution of opacity and decreased optical depth for the rest wavelength (A). A distinct blue-shifted component with $RV$=\,--5\,km$\cdot$s$^{-1}$ (B) arises from a narrow  region (marked with a blue diamond) behind the shock front located at 0.9\,$R_\star$, similar to the previous case of model~W. Moreover, a clear increase of optical depth in the layers in front of the shock (marked with a red parallelogram) can be recognised in the lower right panel for the radial plot corresponding to wavelength point C if velocity effects are taken into account. This leads to an additional red-shifted component at $RV$=\,7.9\,km$\cdot$s$^{-1}$. Altogether, two separated components can clearly be seen in the synthetic spectrum based on this phase of model~S. 

The phenomenon of line doubling is therefore not a matter of the occurrence of shock waves but rather depends on the question if there is infalling material with high enough densities to produce optical depths which are necessary to give rise to a second component in the line profile. By a comparison of Figs.\,\ref{f:massenschalenW}, \ref{f:structureSRV} and \ref{f:co16mueprofiles-WHyavgl} it can be understood that the complex temporal variation of the atmospheric structure of model~W (due to the pulsation) does not lead to a second line component at any time during the light cycle. For $\phi_{\rm bol}$=\textit{0.99} there is only outflow, and a blue-shifted component arises. At the later phase of $\phi_{\rm bol}$=\textit{0.16} the density gradient becomes less steep, and the gas velocities slightly vary around 0\,km$\cdot$s$^{-1}$. As they are formed over a relatively wide range, the spectral line becomes asymmetric, but does not show a pronounced shift. By the time of $\phi_{\rm bol}$=\textit{0.48}, the density structure shows a significant steplike decrease due to the emerging shockwave. Still, all of the infalling layers contribute to the resulting line profile (cf. discussion in Sect.\,\ref{s:realworld} and Fig.\,\ref{f:asymmetry}), which appears red-shifted. After this, the outflowing layers behind the shock become more and more important due to the ever increasing densities. At the same time, the infalling layers in front of the shock loose their influence because of the dramatic decrease in density. At no phase the model shows a clear velocity difference $\Delta u_{\rm gas}$ together with relevant densities in the outer layers. Thus, no line splitting can be found, as exemplified in Fig.\,\ref{f:when-doubling} for the phase $\phi_{\rm bol}$=\textit{0.68}.

\section{Remarks on different velocities}
\label{s:remarkvels}

We note that the derived $\Delta RV$ of the two components of a doubled line profile does not provide information concerning the propagation velocity of the shock wave $u_{\rm front}$, but reflects the difference in velocity $\Delta u_{\rm gas}$ of the outflowing gas behind the shock front and of the infalling matter in front of it. This will be illustrated on the basis of model~M. With the help of the plot of moving mass shells in Fig.\,\ref{f:massenschalenM} the shock propagation velocity\footnote{Note that $u_{\rm front}$ becomes higher than the initial pulsation velocity resulting from the mechanical energy input due to the piston, which is in the case of model~M not larger than 6\,km\,s$^{-1}$ (Table\,\ref{t:dmaparameters}). While the sound waves propagate outwards, they steepen (as described by GH04, see their Fig.\,4.15) and become strong radiating shock waves.} can be estimated to d$R$/d$\phi_{\rm bol}$\,$\approx$\,1.5\,$R_\star$/\,cycle leading to $u_{\rm front}$\,$\approx$\,10.16\,km\,s$^{-1}$. On the other hand, a typical velocity difference of the model at the location of the shock amounts to $\Delta u_{\rm gas}$\,$\approx$\,29\,km\,s$^{-1}$ (cf. Fig.\,\ref{f:structure}), if we take the line doubling phase $\phi_{\rm bol}$=\textit{0.72} (Figs.\,\ref{f:cn2mueprofiles-modelM}+\ref{f:rvsmodelM}) as an example.

However, the most negative values for $RV$ derived from e.g. CO second overtone lines (Fig.\,\ref{f:rv-CO-dv3-modelsMS}) provide a hint on the shock wave velocity $u_{\rm front}$. At least a lower limit can be given. For example, we find $RV$\,$\approx$\,--6.25\,km\,s$^{-1}$ for the blue-shifted component of such a line at phase $\phi_{\rm bol}$=\textit{0.93} of model~M.\footnote{This is the RV value derived from the corresponding line profile at the highest spectral resolution of $R$\,=\,300\,000, rebinning the spectrum to 70\,000 influences the line shape and leads to a slightly different RV compared to those plotted in Fig.\,\ref{f:rv-CO-dv3-modelsMS}.} Multiplying this with a reasonable conversion factor $p$ (Sect.\,\ref{s:factorp}) leads to gas velocities  close to the maximum outflow of --9.6\,km\,s$^{-1}$ behind the shock front (compare the velocity structure for this phase in Fig.\,\ref{f:structure}c with the shock located at $\approx$\,1.25\,$R_\star$), the latter being interrelated with $u_{\rm front}$.

Another type of velocity, which can be studied with the help of our models, is the movement of the formed dust shells. Considering the one recognisable at $\approx$3--5\,$R_{\star}$ in Fig.\,1 of Paper\,I we can estimate  d$R$/d$\phi_{\rm bol}$$\approx$\,1.296 or a shell velocity of $u_{\rm shell}$$\approx$10.5\,km\,s$^{-1}$.

\section{Summary}

The outer layers of evolved red giants during the AGB phase differ strongly from the atmospheres of most other types of stars. Due to pulsations of the stellar interior and the development of a stellar wind, AGB atmospheres are far from hydrostatic equilibrium and the atmospheric structures show strong local variations (spatially and temporally). This leads to a complex line formation process. Many layers at different geometrical depths and with various conditions \mbox{($T,$ $\rho ,$ $u$)} contribute significantly to the finally observable spectra, which are dominated by numerous molecular absorption features. 

Going to high spectral resolutions for studies of individual spectral lines, the complicated velocity fields become especially important. Relative macroscopic motions with gas velocites of the order of 10\,km\,s$^{-1}$ heavily influence line profiles and their evolution in time. Doppler-shifts of spectral features (or components of a line) represent an indicator for kinematics in the atmospheric regions where the lines are formed. Thus, spectroscopic monitoring of molecular lines originating at different atmospheric depths allows us to trace the overall dynamics and explore the process of mass loss in AGB stars.

On the basis of a typical atmospheric structure and by assuming chemical equilibrium as well as LTE, we studied the influence of the parameters ($E_{\rm exc}$ and $gf$) of spectral features on the respective estimated line-forming region within the atmosphere (Sect.\,\ref{s:lineformation}). This is of particular importance for the different rotation-vibration bands of CO, which play an important role in the course of kinematics studies. 

We proceeded with the second aspect of this paper, namely the modelling of line profiles with the help of state-of-the-art dynamic model atmospheres. For this purpose, we used models which represent a numerical simulation of the most widely accepted (at least for C-rich stars) scenario of pulsation-enhanced dust-driven winds. Atmospheric structures at certain instances of time were used as input for the spectral synthesis, for which we assumed conditions of LTE as well as equilibrium chemistry. A spherical RT code was applied that accounts for velocity effects.

Spectroscopic observations of Mira variables revealed quite characteristic line profile variations over the light cycle for spectral features probing the inner dust-free atmospheric layers where the atmospheric movements are ruled by the pulsating stellar interiors. We were able to tune the input parameters of a dynamic model (model~M) so that we could quantitatively reproduce the -- apparently universal -- discontinuous RV curve (line doubling at phases of light maximum, S-shape, amplitude, asymmetry) of CO $\Delta v$\,=\,3 (and CN) lines, which can be interpreted as a shock wave propagating through the region of line formation (Sect.\,\ref{s:largerampl}).
To match observed RV curves of these lines with synthetic ones based on model~M we needed to introduce a phase shift of $\Delta\phi$\,$\approx$\,0.3 between bolometric phases $\phi_{\rm bol}$ and visual ones $\phi_{\rm v}$ so that the former would lag behind the latter (in agreement with photometric results).

In observed spectra of SRVs the same spectral lines show a different behaviour. Although there seems to be no general RV curve for this type of stars (as for Miras), a RV distribution peaking around phases of light minimum and no line doubling is commonly found. Synthetic line profile variations (Sect.\,\ref{s:COdv3modelW}) computed on the basis of a pulsating model atmosphere (model~W; no wind) are able to reproduce this and compare well to observational results of the semiregular variable W\,Hya (showing an almost negligible mass loss of $\dot M$\,$\approx$\,2$\cdot$10$^{-8}$\,$M_{\odot}$\,yr$^{-1}$). The same model was used to investigate the phenomenon of line doubling (Sect.\,\ref{s:when-linedoubling}). We argue that a shock wave propagating through the line forming region may not be sufficient for the occurence of doubled lines in every case. Large enough optical depths in the infalling layers may be necessary for a second, red-shifted component to be visible in the spectra. This could be an explanation for the fact that for some stars we find observational evidences for shock waves in their atmospheres (emission lines), but do not observe doubled CO~$\Delta v$=3 lines.

The connection between gas velocities in the line-forming region and RVs derived from Doppler-shifted spectral lines was examined by means of artificial velocity fields (Sect.\,\ref{s:factorp}). On the basis of a hydrostatic initial model and under the assumption of a constant value for $u_{\rm gas}$ at every depth point of the atmosphere we found conversion factors $p$ of \mbox{$\approx$1.2--1.5.} We also point out possible difficulties for relating measured RVs with real gas velocities (e.g. inferring the velocity difference across a shock front from doubled lines) due to the complicated line formation process.

In summary the models studied in this work represent the outer layers of pulsating AGB stars with or without mass loss quite realistically. At least for the C-rich case, the dynamic model atmospheres approach quantitative agreement with observations. One important aspect in the comparison of modelling results with observational findings are line profile variations in high-resolution spectra. Agreement in the velocity variations of various molecular NIR features sampling different regions within AGB atmospheres can provide information on the interrelation of pulsation and mass loss or set constraints on the acting mass loss mechanism. In addition, this may be a criterion to constrain stellar parameters of a model. 

Although selected observed dynamic aspects (CO $\Delta v$\,=\,3 lines in Mira spectra) are reproduced very well by the atmospheric models and also the global velocity field resembles what can be derived from observations of typical long period variables, a fine-tuning of the model parameters to fit certain objects appears to be more difficult due to the complex dependence on several parameters.

In contrast to the encouraging results for the second overtone lines of CO in both Miras and SRVs, the case of CO first overtone lines is a lot less clear-cut both from the observational and the theoretical point of view. These lines are presumably formed within the dust formation and wind acceleration region, and they display quite different types of behaviour in the few well-observed stars, with a considerable spread in variability. A similarly varied picture -- probably an indication of the dynamical complexity of the corresponding layers -- emerges from the models that we have analysed in detail here and in earlier papers. In particular, model~M (which reproduces the second overtone lines nicely) shows basically no variations of the first overtone lines. We are, however, confident that further tuning of stellar parameters can produce a model which shows the typical behavior observed for second overtone lines and, at the same time, more pronounced variations in the first overtone lines.

A detailed comparison of phase-dependent line profiles derived from models with observations is presently limited by uncertainties in the relation between bolometric and visual lightcurves. Bolometric phases -- which are quite directly coupled to global atmospheric dynamics (pulsation and resulting shocks) -- are a natural choice for discussing model features. Observations, on the other hand, usually are given in terms of visual phases, as bolometric variations are not directly accessible. In principle, visual phases can be derived for the models, but they may depend strongly on details of dust processes (grain opacities can dominate in the visual range for C-stars, cf. Appendix\,\ref{s:phaseshift}), which may have only a weak connection with the dynamical phenomena we want to probe by studying the line profiles. A more comprehensive discussion of photometric model properties and their observed counterparts will be the topic of a separate paper.
\newline

\begin{acknowledgements}
The observational results (RV data as well as FTS spectra of $\chi$\,Cyg, S\,Cep, and W\,Hya) were kindly provided by Th. Lebzelter and K. Hinkle. Sincere thanks are given to Th. Lebzelter and J. Hron for careful reading and fruitful discussions. This work was supported by the \textit{Fonds zur F\"orderung der Wis\-sen\-schaft\-li\-chen For\-schung} (FWF) under project numbers P18939--N16 and P19503--N16 as well as the Swedish Research Council. BA acknowledges funding by the contract ASI-INAF I/016/07/0. We thank the referee, Peter Woitke, for several valuable comments.
\end{acknowledgements}

\appendix
\section{Bolometric vs. visual light variations}
\label{s:phaseshift}

\subsection{Observational findings}
\label{s:shiftobs}

A direct comparison of observed and synthetic line profiles (or derived RVs) is affected because different phase informations are used in the two cases. On the one hand, observations are usually linked to visual phases $\phi_{\rm v}$ within the lightcycle (e.g. AAVSO measurements). On the other hand, for the model calculations -- a priori -- only bolometric phases from the luminosity lightcurve (e.g. Fig.\,\ref{f:structureSRV}) are known, which is itself determined by the inner boundary (cf. Sect.\,\ref{s:DMAsused}). A relation between the two kinds of phase information would be desirable for comparing the modelling results with observational findings.

Measuring bolometric lightcurves of LPVs is not straigthforward. Although a few authors estimated such variations from multi-epoch observations in several filters by fitting blackbody curves (e.g. Whitelock et al. \cite{WhiMF00}), a bolometric lightcurve is in general not available for a given star. However, it is known that lightcurves in the NIR (e.g. the photometric K-band) trace bolometric variations more closely than lightcurves in the optical wavelength range. The reasons for this are that (i) most of the flux is emitted in the NIR, and (ii) the visual spectra are severely influenced by features of temperature sensitive molecules.\footnote{The latter effect is especially pronounced in the O-rich case and the characteristic bands of the fragile TiO molecule. It may be of less importance in the C-rich case as the species responsible for features in the visual, CN and C$_2$, are more stable.} Infrared lightcurves may therefore provide clues on the desired \mbox{$\phi_{\rm bol}$-\,$\phi_{\rm v}$\,-\,relation.} 

Several studies compared variations in the visual and NIR of Mira variables in the past (cf. the list of references given by Smith et al. \cite{SmiPM05}), leading to a rather general result: \textit{the maxima of IR (i.e. $\sim$bolometric) lightcurves lag behind the maxima of visual lightcurves by $\approx$0.1--0.2 in phase.} A good illustration of this is provided e.g. by Fig.\,2.44 of Lattanzio \& Wood (\cite{LattW04}), who show lightcurves of the \mbox{M-type} Mira RR\,Sco in various photometric bands. By using a well chosen narrow-band filter at 1.04\,$\mu$m (located in a wavelength region with almost no absorption features), Lockwood \& Wing (\cite{LockW71}) and Lockwood (\cite{Lockw72}) also found a typical shift of $\Delta\phi$$\approx$0.1 if the phase values measured by this filter are assumed to correspond to bolometric ones.

Recently, Smith et al. (\cite{SmiPM05}) compared IR photometric data (1.25--25$\mu$m) of the DIRBE instrument onboard the COBE satellite with visual observations (mainly AAVSO) for a sample of 16 Miras and 5 SRVs. Their results confirmed the phase lag of 0.1--0.2 between visual and IR phases for Miras. In contrast, they found no phase lags for SRVs. Smith et al. could reproduce both results also with synthetic spectra based on dynamical model atmospheres of the Australia-Heidelberg collaboration (O-rich, dust-free). They conclude that phase lags are in general more likely or larger for more evolved stars (higher $L$, lower $T_{\rm eff}$, higher $\dot M$, higher $\Delta m_{\rm bol}$, larger periods), that is for Miras rather than for SRVs.

Most of the investigations in this context were carried out for M-type stars, results for \textit{C-rich LPVs} are rather scarce in the literature. Thus, \textit{only some indications} can be listed here. Kerschbaum et al. (\cite{KersLL01}; their Fig.\,5) found that their lightcurve in the K-filter of the C-rich Mira T\,Dra lags behind the visual data (AAVSO) by 0.06 in phase. Smith et al. (\cite{SmiPM05}) state that they find no differences in phase lag for the spectral types M/S/C. For the one C-type Mira in their sample, V\,CrB, Smith et al. derive shifts of $\Delta\phi$=0.13--0.15, the C-rich SRV UX\,Dra does not show any shift. Smith et al. (\cite{SmiPM05}) discussed optical-IR-offsets of broadband colours computed on the basis of the same type of C-rich dynamic model atmospheres as used here. They report that some of the models show phase lags, but do not give numbers. However, no clear trend in the sign of the phase shift was found. For some models, visual phases precede IR ones, while it is the other way round for other models.

\subsection{Modelling results}
\label{s:shiftmod}

\begin{figure}
\resizebox{\hsize}{!}{\includegraphics[clip]{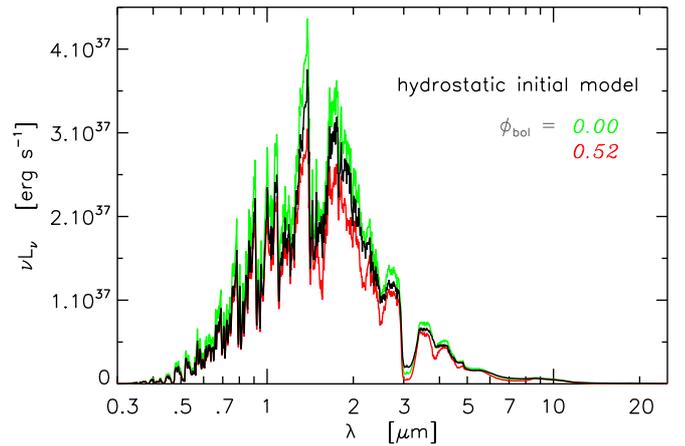}}
\caption{Low-resolution opacity sampling spectra based on the initial model as well as selected phases of model~W (dust-free, no wind).} 
\label{f:specvarmodw}
\end{figure}

\begin{figure}
\resizebox{\hsize}{!}{\includegraphics[clip]{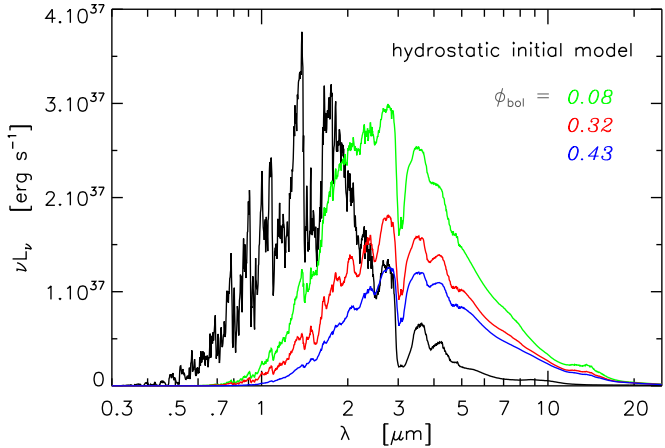}}
\caption{Same as Fig.\,\ref{f:specvarmodw} but for model~M (dusty wind).}
\label{f:specvarmodm}
\end{figure}

Following the suggestions of the referee, we studied the photometric variations in the visual of the models used in this paper. Based on the dynamic model atmospheres of Tab.\,\ref{t:dmaparameters} and consistent with the descriptions in Sect.\,\ref{s:synthesis}, we calculated low-resolution opacity sampling spectra (cf. Gautschy-Loidl et al. \cite{GaHJH04}) with extensive wavelength coverage. By convolving these with transmission curves (Bessell \& Brett \cite{BB88}, Bessell \cite{Besse90}) for filters of the standard Cousins-Glass-Johnson system and applying adequate zeropoints we get synthetic photometry. For details of the method we refer to Aringer et al. (\cite{AGNML09}).

Example spectra (rebinned to $R$\,=\,360) are shown in Figs.\,\ref{f:specvarmodw}+\ref{f:specvarmodm}, demonstrating the qualitative difference between a model for a moderately pulsating red giant on the one hand and an evolved Mira with dust formation and mass loss on the other hand. In the first case, the spectrum shows only minor variations around the hydrostatic case during the pulsation cycle. The significantly changed radial structure in the latter case (Sect.\,\ref{s:DMAsused}) is also reflected in the spectra based on wind models. Due to the characteristic absorption behaviour of dust grains composed of amorphous carbon (e.g. Fig.\,1 of Andersen et al. \cite{AndLH99}), the whole spectral energy distribution is shifted towards longer wavelengths. It does not resemble the hydrostatic spectrum at any point in time.

\begin{figure}
\resizebox{\hsize}{!}{\includegraphics[clip]{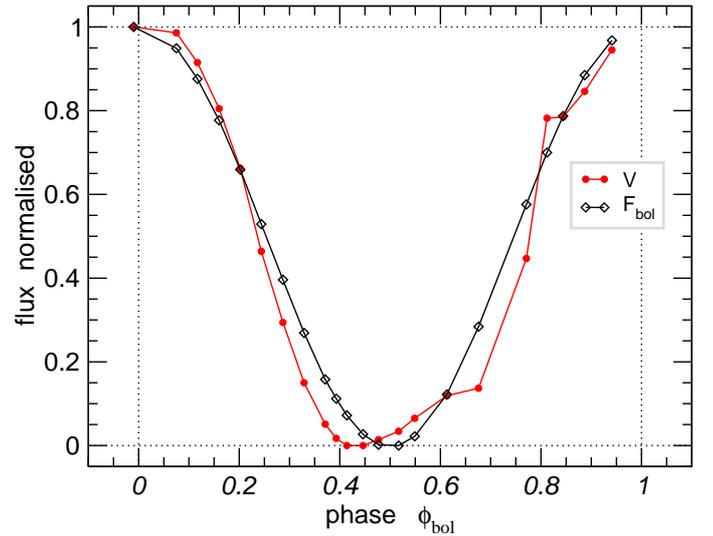}}
\caption{Bolometric variation of model~W (caused by the variable inner boundary; Sect.\,\ref{s:DMAsused}) during one pulsation cycle and the resulting synthetic lightcurve in the visual.}
\label{f:lightcurvemodw}
\end{figure}

For an easier comparison the synthetic lightcurves in various filters are plotted with fluxes normalised to [0,1] below.\footnote{Note that the amplitude of the variations may differ strongly between filters, with amplitudes decreasing with increasing wavelengths.} As shown in Fig.\,\ref{f:lightcurvemodw}, the light variation in the V-filter of the moderately pulsating model~W follows the bolometric lightcurve quite closely, although some slight deviations from the pure sinusoidal shape are apparent.

\begin{figure}
\resizebox{\hsize}{!}{\includegraphics[clip]{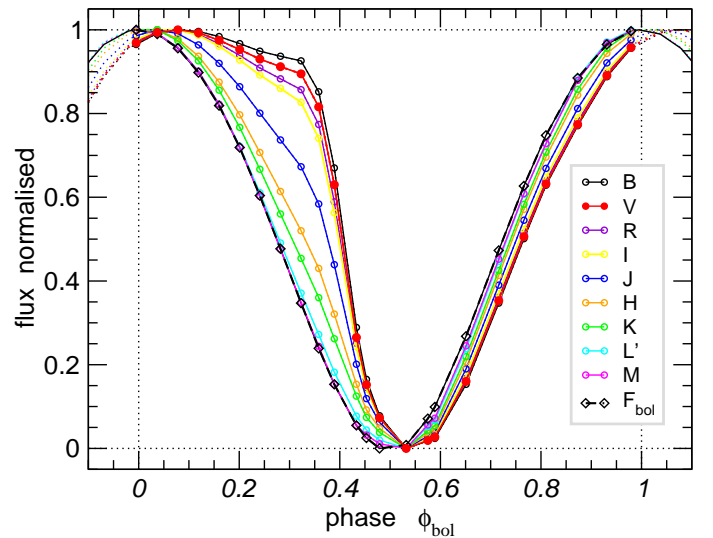}}
\caption{Synthetic normalised lightcurves for model~M for different filters throughout one pulsation cycle.}
\label{f:lightcurvemodm}
\end{figure}

Figure\,\ref{f:lightcurvemodm} shows the light variations in various filters for the model~M with mass loss. As found in observational studies, IR-lightcurves stay rather close to the bolometric one. The lightcurves in the visual and red show some deviation, though. A sinusoidal behaviour with a phase lag of $<$\,0.1 is found for most phases throughout the lightcycle (ascending branch), with visual lightcurves lagging behind the IR/bolometric ones, in contrast to observational results. Significant differences between filters are seen on the descending branch of the lightcurve. The more or less pronounced asymmetries -- monotonically increasing with decreasing wavelengths -- suggest a relation with the absorption by amorphous carbon dust.

\begin{figure}
\resizebox{\hsize}{!}{\includegraphics[clip]{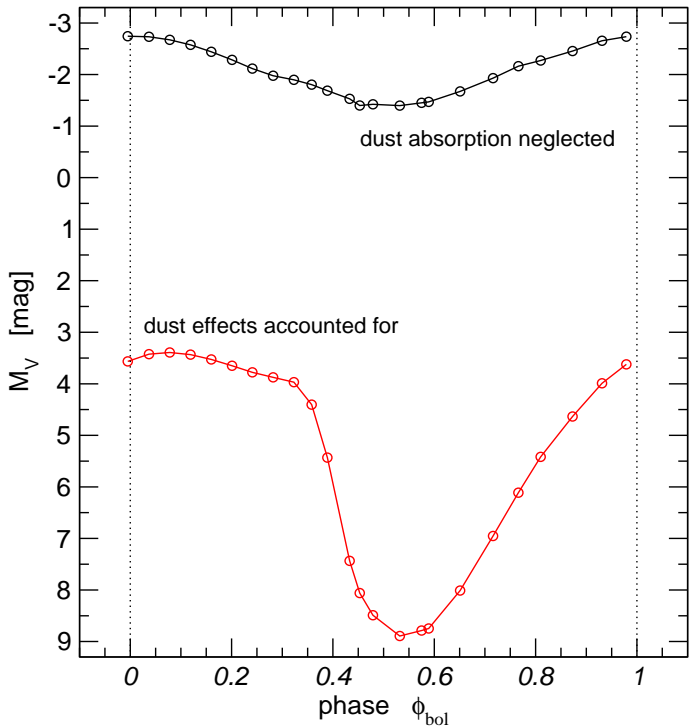}}
\caption{Synthetic lightcurves (absolute magnitudes) based on
model~M in the V-filter, where the absorption of circumstellar
carbon dust is taken into account (same as
Fig.\,\ref{f:lightcurvemodm}) or not for the spectral synthesis.}
\label{f:lightcurvemodmdusteffect}
\end{figure}

Figure\,\ref{f:lightcurvemodmdusteffect} confirms this idea. If the dust absorption is not taken into account for computing spectra and photometry (not consistent with the dynamical modelling but a useful test), no asymmetry and phase shift is found for the \mbox{V-lightcurve.} If the dust absorption is included, the V-band flux of the star becomes considerably fainter and the asymmetry of the visual lightcurve becomes apparent. The pronounced decrease in $V$ as seen in Fig.\,\ref{f:lightcurvemodmdusteffect} is related to the new dust shell emerging just before light minimum for model~M (see Fig.\,\ref{f:massenschalenM}). The increasing dust absorption causes deviations from a sinusoidal variation and shifts the time of minimum (and maximum) light towards higher values of $\phi_{\rm bol}$. This may provide an explanation for why a phase discrepancy of $\approx$\,0.3 is found in Sect.\,\ref{s:largerampl} (Mira observations vs. the dusty model~M), while only a value of $\approx$\,0.1 is needed in Sect.\,\ref{s:COdv3modelW} (SRV observations vs. the dust-free model~W). Assigning one global value for $\Delta\phi$\,=\,$\phi_{\rm bol}$\,--\,$\phi_{\rm v}$ is impeded by these deviations from a sinusoidal behaviour.

Considering the possibly dominant role of dust opacities at visual wavelengths, however, any conclusions drawn from comparing observed and synthetic light curves (phases) should be regarded with caution for the current models. Recent modelling results indicate that grains in AGB star winds may grow to sizes where the small particle limit (used for the dynamical calculations here) is no longer a good approximation for dust opacities, in particular at wavelengths shorter than about 1\,micron (cf. Mattsson et al. \cite{MatWH09}, Mattsson \& H\"ofner in prep.). Unfortunately, model~M is no exception in that respect, with a mean grain size of about $10^{-5}$\,cm. For these grains, both the absolute and relative values of opacities at different wavelengths become dependent on the size of the particles. This may have a significant impact on photometric light curves. Phase-dependent effects due to grain growth may occur in different filters, possibly affecting the relative positions of maxima and minima. A detailed investigation of this effect is, however, beyond the scope of this paper and will be discussed in a future paper about synthetic photometry.

In summary, at the current state of modelling, bolometric light curves can be regarded as a quite reliable indicator of the atmospheric dynamics of the models, since they are closely linked to the effects of pulsation and atmospheric shocks. In contrast, the diagnostic value of the synthetic V-lightcurves -- and, consequently, of the synthetic visual phases -- is rather doubtful for this purpose as the fluxes in the visual are strongly dependent on grain opacities (which probably require improvements in the future) and time-dependent dust processes, which may only have a weak connection with atmospheric dynamics. Therefore, with the additional uncertainties concerning the \mbox{$\phi_{\rm bol}$-\,$\phi_{\rm v}$\,-\,relation} on the observational side (as outlined in the previous subsection), we decided to use the more representative bolometric phases when discussing modelling results in this paper.

\end{document}